\documentstyle[12pt,axodraw,psfig,epsf]{article}

\newcommand{\eq}[1]{(\ref{#1})}
\newcommand{\ab}{( \alpha - \beta )} 
\newcommand{\nn}{\nonumber}
\newcommand{\fr}{\frac}
\newcommand{\mh}{m_{h^0}^2}
\newcommand{\mH}{m_{H^0}^2}
\newcommand{\mg}{m_{H^{\pm}}^2}
\newcommand{\ma}{m_{A^0}^2}

\newcommand{\mw}{m_W^2}
\newcommand{\mz}{m_Z^2}

\newcommand{\al}{\alpha}
\newcommand{\be}{\beta}
\newcommand{\sn}{\sin}
\newcommand{\cs}{\cos}

\newcommand{\tnb}{\tan \beta}
\newcommand{\ctb}{\cot \beta}
\newcommand{\snab}{\sin (\alpha - \beta)}
\newcommand{\csab}{\cos (\alpha - \beta)}

\newcommand{\lt}{\left}
\newcommand{\rt}{\right}

\newcommand{\gmn}{g_{\mu\nu}}
\begin{document}
\topmargin 0pt
\oddsidemargin 1mm
\begin{titlepage}
\begin{flushright}
 KEK Preprint 97-160\\
 KEK-TH- 544\\
\today
\end{flushright}

\setcounter{page}{0}
\vspace{10 mm}
\begin{center}
{\Large \bf   Enhancement of Loop Induced    
                $H^\pm W^\mp Z^0$ Vertex
                in Two Higgs-doublet Model   }
\end{center} 
\vspace{15 mm}
\begin{center}
{\large \bf  Shinya Kanemura 
  \footnote{e-mail: kanemu@theory.kek.jp}}\\    
\vspace{2mm}
{\em Theory Group, KEK,\\
       Tsukuba, Ibaraki 305, Japan}\\
\end{center}

\vspace{15 mm}

\begin{abstract}
The non-decoupling effects of heavy Higgs bosons as well as fermions 
on the loop-induced $H^\pm W^\mp Z^0$ vertex are discussed in the 
general two Higgs doublet model.  
The decay width of the process $H^+ \rightarrow W^+ Z^0$ 
is calculated at one-loop level and the possibility of its enhancement 
is explored both analytically and numerically. 
We find that the novel enhancement of the decay width can be realized 
by the Higgs non-decoupling effects with large mass-splitting 
between the charged Higgs boson and the CP-odd one.     
This is due to the large breakdown of the custodial $SU(2)_V$ 
invariance in the Higgs sector.    
The branching ratio can amount to 
$10^{-2} \sim 10^{-1}$ for $m_{H^\pm} = 300$ GeV 
within the constraint from the present experimental data. 
Hence this mode may be detectable 
at LHC or future $e^+e^-$ linear colliders.   
\end{abstract}

\vspace{12mm}

\end{titlepage}

\section{Introduction}

\hspace*{12pt}
The standard model (SM) of the electroweak interaction has been tested 
by a lot of experiments and any substantial deviation from the data from 
the precision experiments such as at LEP and SLC \cite{lep} 
has not been found so far \cite{hmh}. 
In spite of the success of SM, the symmetry breaking sector (Higgs sector) 
remains unknown. The Higgs sector is expected to be probed at 
LEP II \cite{lep2},  LHC \cite{lhc} and future $e^+e^-$ linear colliders 
(LC's) \cite{lc}. 
Although the minimal Higgs sector with only one Higgs doublet is consistent 
with the available experimental data, the Higgs sector may be favored to 
have rather a little more complicated structures from the various 
theoretical viewpoint \cite{hhg}. 
One of the simplest but rich extensions of the minimal Higgs sector is 
the two Higgs doublet model (THDM). 
There are a lot of motivations for THDM such as the minimal 
supersymmetric standard model (MSSM), 
additional CP violating phases, 
and a solution of the strong CP problem.

The charged Higgs boson $H^{\pm}$ is one of the new particle contents 
in such the extended Higgs sectors and its exploration is a quite important 
task of the future experiments at LHC and LC's.
If $H^{\pm}$ is relatively light ($ < m_W$), it may be detected at LEP II.
If $H^\pm$ has the mass of the intermediate scale 
($m_W < m_{H^\pm} < m_t + m_b$), 
$H^\pm$ would be detected at future colliders such as LHC or 
LC through the decay modes $H^\pm \rightarrow \tau \nu$ and $c s$. 
Alternatively if $m_{H^\pm}$ is large enough to allow 
$H^\pm \rightarrow t b$ kinematically, 
it seems to be difficult to detect it because of the large QCD background.  
The heavy $H^\pm$ is, in fact, favored by the results of 
the $b \rightarrow s \gamma$ measurement 
in THDM with Type II Yukawa couplings \cite{cleo}
(There are narrow loopholes for this constraint in MSSM \cite{goto}.).     
In such cases, we have to investigate the possibility of the alternative modes 
with the branching ratio enough to yield substantial events to probe $H^\pm$. 
The possible modes for such the purpose may be $H^\pm \rightarrow \tau \nu$, 
$h^0 W^\pm$, $W^\pm Z^0$ and $W^\pm \gamma$. 
Unfortunately, it has been known that the latter two modes 
(namely, decays into a gauge boson pair) disappear at tree level in general 
multi Higgs-doublet models including THDM.  
This property is quite different from the case of (CP-even) neutral 
Higgs bosons. 
The absence of the tree $H^\pm W^\mp \gamma$ coupling comes from the current 
conservation of $U(1)_{\rm em}$. 
On the other hand, tree $H^\pm W^\mp Z^0$ coupling is absent because of the 
isospin symmetry of the kinetic term of the Higgs sector \cite{glme}.       
Since both these characteristics are, in general, broken at one-loop level 
through effects from other sectors, 
these vertices are induced at loop level. 
Hence the question of how large the vertex can  
be enhanced by loop effects occurs.
In fact, the estimations of the loop-induced decay widths for 
$H^\pm \rightarrow W^\pm Z^0$ and $H^\pm \rightarrow W^\pm \gamma$ 
have been studied in part by several authors: 
Pomarol and Mendez \cite{mepo} 
once have studied $H^+ \rightarrow W^+ Z^0$ in MSSM, 
and Capdequi Peyranere et al. \cite{capd} 
have calculated the fermion and sfermion non-decoupling effects on 
$H^+ \rightarrow W^+ Z^0$ and $H^+ \rightarrow W^+ \gamma$. 
In their works, it has been pointed out that 
while the loop-induced $H^\pm W^\mp \gamma$ vertex is much small, 
the loop-induced $H^\pm W^\mp Z^0$ vertex can be largely enhanced by 
the mass effects of {\it super heavy} fermions.
It, however, seems to be difficult to consider such the heavy fermions in the 
present situation that the top-quark has already 
been discovered at $\sim 175$ 
GeV \cite{top} and that the fourth generation of fermions has been almost 
excluded by the S-parameter constraint \cite{hmh,4gen}.    
Hence the substantial enhancement of the vertices seems no longer to  
be possible in the framework of MSSM, in which the non-decoupling effects 
on the vertex are essentially only due to heavy fermions.
Apart from MSSM, there are some exotic Higgs models with 
more complicated Higgs multiplets (for example, triplets)  
which have $H^\pm W^\mp Z^0$ coupling at tree level \cite{exo}.   
Therefore the decay mode $H^+ \rightarrow W^+Z^0$ has often been 
considered as a clear signature for these exotic Higgs sectors.

Here occurs another question of how about the general THDM but MSSM.    
In MSSM, the Higgs mass-effects are very small because of their decoupling 
property and the fermion effects are dominant.    
On the other hand, there can be non-decoupling effects of heavy Higgs bosons 
in THDM in general. 
In such the model, 
whether the loop induced $H^\pm W^\mp Z^0$ vertex can be substantially 
enhanced by the Higgs mass-effects or not is a non-trivial and 
also very interesting problem.
In the previous works \cite{mepo,capd,riz}, 
these effects have not been considered at all.
The purpose of this paper is just to solve this problem.

In this paper, we discuss the loop induced $H^\pm W^\mp Z^0$ vertex in THDM.   
The possibility that this vertex can be largely enhanced 
by the non-decoupling effects of the heavy Higgs bosons is explored 
in detail.
In the THDM Higgs sector, whether the heavy Higgs bosons are decoupled or not 
is rather model dependent.
The heavy Higgs bosons, in general, 
receive their masses from both the vacuum expectation 
value and the (bare) non-zero soft-breaking parameter. 
If the contribution of the latter effect is relatively dominant, 
the masses become approximately independent of 
the self-coupling constants and then the heavy Higgs bosons become decoupled 
\cite{dec}. 
The MSSM Higgs sector approximately corresponds to this case, in which  
all the quartic self-coupling constants are constrained to 
$\sim {\cal O}(g^2)$ and the Higgs bosons other than the lightest can 
become heavy only by the growing soft-breaking parameter. 
The effects of these heavy masses then should be suppressed by the 
decoupling theorem. 
This is one of the main reasons why the Higgs mass effects are less important 
than the fernion's ones in MSSM. 
Alternatively, if the heavy Higgs bosons are due to the large 
self-coupling constants, 
the masses naively become proportional to the coupling 
constants and then the non-decoupling effects of the Higgs bosons 
can be expected as well as those of the fermions \cite{ab,nondec,kt}. 
Such the non-decoupling effects of the heavy Higgs bosons
(for example, the effects of the heavier neutral boson
$H^0$ or the CP-odd Higgs boson $A^0$) are of our central interest here.

From the naive power counting, the non-decoupling effects on 
the $H^\pm W^\mp Z^0$ vertex can be expected to include quadratic 
and logarithmic mass contributions at one loop level \cite{capd}. 
\footnote{In the case of $H^\pm W^\mp \gamma$ vertex, 
only the logarithmic mass effects are possible because of the $U(1)_{em}$ 
current conservation \cite{capd}. 
Thus the non-decoupling effects cannot be so large. 
Hence we here consider $H^\pm W^\mp Z^0$ vertex only.}
However, by making the effective Lagrangian it is shown that 
these non-decoupling effects are completely canceled 
if the theory has the global custodial $SU(2)_V$ symmetry.   
There is the similar situation in the oblique corrections 
known as the screening theorem \cite{sc}.  
In THDM, the custodial symmetry in the Higgs sector is 
explicitly broken except for the case of the mass degeneracy between 
the $H^\pm$ and $A^0$ \cite{haber,kkt}.  
Hence we expect that the large mass splitting leads to the quadratic 
Higgs mass effects to the vertex. 
Thus the conditions for large enhancement of the vertex by 
the heavy Higgs bosons are 
1) the large Higgs masses coming from the larger 
   contributions of the quartic coupling constants 
   with keeping the soft-breaking parameter to be smaller, 
2) the large explicit breaking of $SU(2)_V$ in the Higgs sector 
   by large mass splitting between $H^\pm$ and $A^0$.    
We note that although THDM is indeed strongly constrained by the experimental 
results for the oblique corrections \cite{gr}, 
there remains large allowed region for  
large mass splitting between $H^\pm$ and $A^0$ because of 
a lot of free parameters in the model \cite{kt}.

With the consideration above, we analyze the decay process 
$H^+ \rightarrow W^+ Z^0$ at one loop level in THDM. 
The calculation is performed in the t'Hooft-Feynman gauge
for the Higgs-Goldstone sector and gauge sector and
in the unitary gauge for fermion loops.
Since all the diagrams with a fermion-loop themselves
construct a gauge invariant subset, we are free to use
different gauge choices like above \cite{capd}.
In addition to the conditions above, the vertex turns out to be 
much sensitive to the Higgs mixing angles.  
We find that the branching ratio $Br(H^+ \rightarrow W^+ Z^0)$ 
at $m_{H^\pm} = 300$ GeV  
become larger than $10^{-2}$ if the mass splitting between 
$H^+$ and $A^0$ is larger than 200 GeV for $\tan \be > 6 \sim 8$, 
where $\tan \be = v_2/v_1$.  
The maximal value of $Br(H^+ \rightarrow W^+ Z^0)$ 
can amount to near $10^{-1}$ for $\tan \be > 20$ and 
very large $m_{A^0}$ but within the allowed region 
from the tree-level unitarity bound \cite{uni}. 
Such the enhancement of the branching ratio 
($10^{-2} \sim 10^{-1}$) may make it possible to detect the mode at LHC 
\cite{lhc}.
It is expected that more than a few dozen of the events 
($H^\pm \rightarrow W^\pm Z^0 \rightarrow lll\nu$) 
are produced for the branching ratios of $\sim 10^{-2}$.  
As to the background (mainly from $u \bar{d} \rightarrow W^+Z^0$), it is 
likely such that the branching ratio of a few \% would be required 
in order to see a signal \cite{back}.  
Therefore in THDM 
the non-decoupling effects of Higgs bosons can induce a significant
enhancement of the branching ratio and this process may become 
detectable at LHC. 
In the $SU(2)_V$ symmetric cases ($m_{A^0} \sim m_{H^\pm}$), 
the Higgs non-decoupling effects are canceled out and only the fermion 
and gauge boson contributions remain, 
so that the branching ratio becomes smaller than $10^{-4}$ for $\tan \be > 1$. 
We also show that such the enhancement is reduced by taking account of 
the soft-breaking parameter to be large.

In Sec 2, we introduce THDM and discuss its decoupling and non-decoupling 
properties. 
Sec 3 is devoted to the qualitative study of the possibility of enhancement
of the $H^\pm W^\mp Z^0$ vertex due to the non-decoupling effects.  
In Sec 4, the decay process $H^+ \rightarrow W^+ Z^0$ is evaluated in THDM 
and the novel enhancement of the branching ratio is shown. 
In Sec 5, we summerize the results and discuss some 
phenomenological implication.  
The explicit results of the calculations are attached in Appendices.

\newpage

\section{The Model}

\hspace*{18pt}
In this paper, we discuss the non-decoupling effects of 
the additional heavy Higgs bosons as well as the fermions 
on the $H^\pm W^\mp Z^0$ vertex in the two Higgs doublet model (THDM). 
We here consider the model with a softly-broken discrete symmetry 
under $\Phi_1 \rightarrow \Phi_1$, $\Phi_2 \rightarrow - \Phi_2$ 
because there are too many parameters to be analyzed in the most general 
THDM. 
The discrete symmetry has often been imposed  
for natural avoiding of the flavor changing neutral current 
(FCNC) \cite{glwe}.  
There are two types of Yukawa sector under the discrete symmetry 
according to the assignment of the charge of quarks, 
what we call, Type-I and Type-II in Ref \cite{hhg}.   
We employ the Type-II coupling in our later calculation.   
The Higgs sector is defined by        
\begin{eqnarray}
  {\cal L}_{\rm THDM}^{int} 
                       & = &  \mu_1^2 \left| \Phi_1 \right|^2 + 
                              \mu_2^2 \left| \Phi_2 \right|^2 + 
                             \lt\{ \mu_{3}^2 
                                  \lt( \Phi_1^{\dagger} \Phi_2 \rt) 
                                     + {\rm \,h.c. \,} \rt\}     \nn  \\
                       &   & - \eta_1 \left| \Phi_1 \right|^4
                             - \eta_2 \left| \Phi_2 \right|^4
                             - \eta_3 \left| \Phi_1 \right|^2 
                                \left| \Phi_2 \right|^2 \nn \\
                       &   & - \eta_4 \left\{ 
                                     \lt( {\rm Re }\Phi_1^{\dagger}
                                                  \Phi_2 \rt)
                                                          \right\}^2
                             - \eta_5 \left\{
                                     \lt( {\rm Im }\Phi_1^{\dagger}
                                                  \Phi_2 \rt)
                                                          \right\}^2. 
\label{int}
\end{eqnarray}
This potential covers the MSSM Higgs sector as a special case \cite{hhg}. 
The soft-breaking parameter $\mu_3^2$ is in general a complex quantity, 
which can give an additional CP violating source \cite{cp}. 
We here confine ourselves in the CP invariant world 
by assuming $\mu_3^2$ to be real in order to 
reduce the number of parameters and also concentrate into 
extracting the essential contribution of the non-decoupling effects 
of the Higgs bosons.  

The Higgs potential \eq{int} has a global symmetry, 
that is, the custodial $SU(2)_V$ symmetry if $\eta_5$ is zero 
\cite{haber,kkt}.
To see this, it is convenient to rewrite RHS in \eq{int} in terms of 
$2 \times 2$ matrices 
${\cal M}_i = \left(  i \tau_2 \Phi_i^{\ast}, \Phi_i \right)$. 
All the terms except for the $\eta_5$-term can be rewritten 
as combinations of 
${\rm tr}({\cal M}_i^{\dagger}{\cal M}_j)$, 
            ($i, j = 1\;{\rm and/or}\; 2$).  
Thereby it becomes clear that 
these terms are invariant under the transformation  
${\cal M}_i \rightarrow g_L^\dagger {\cal M}_i g_R$,     
($g_{L,R} \in SU(2)_{L,R}$, $SU(2)_L$ is the gauge symmetry of the weak 
 interaction.).          Hence, if $\eta_5$ is zero, 
there remains the global symmetry in the Higgs sector
even after the gauge symmetry breaking; 
$SU(2)_L \otimes SU(2)_R \rightarrow SU(2)_V$ \cite{cus}.   
On the other hand,  the $\eta_5$-term is rewritten as 
$\sim \eta_5 
 \left\{{\rm tr}({\cal M}_2 \tau_3 {\cal M}_1^{\dagger}) \right\}^2$.
The $\eta_5$-term breaks $SU(2)_R$ and thus $SU(2)_V$ explicitly.
Since the SM Higgs sector with one doublet is known to $SU(2)_V$ symmetric, 
the explicit breaking of $SU(2)_V$ in the Higgs sector leads to 
a new physics by itself.   
The custodial symmetry plays a crucial role in our later discussion. 
\footnote{As we mention later, 
 the severe constraint of the $\rho$-parameter from the present data 
 do not always forbid the large breaking of the custodial symmetry 
 in the THDM Higgs sector  completely. }

The Higgs doublets both with $Y = 1/2$ are parametrized as 
\begin{eqnarray}
 \Phi_i = \left( \begin{array}{c}
                w^+_i                                              \\
                 \frac{1}{\sqrt{2}} ( h_i +v_i + i z_i )
                   \end{array} 
                   \right) \,,\,\, ( i = 1,2 ), \label{config}
\end{eqnarray}   
where the vacuum expectation values $v_1$ and $v_2$ are combined 
to give $v = \sqrt{v_1^2 + v_2^2} \sim 246{\rm GeV}$.
The diagonalization of the mass matrices is performed by introducing two
mixing angles $\al$ and $\be$ in the following way; 
\begin{eqnarray}
\left( \begin{array}{c}
              h_1                                  \\
              h_2 
        \end{array}  \right)
        =  R(\alpha)
               \left( \begin{array}{c}
                       H^0                           \\
                       h^0 
                        \end{array}  \right),     
\left( \begin{array}{c}
              w^{\pm}_1                                  \\
              w^{\pm}_2 
        \end{array}  \right)
        =   R(\beta)
               \left( \begin{array}{c}
                       w^{\pm}                           \\
                       H^{\pm} 
                        \end{array}  \right), 
\left( \begin{array}{c}
              z_1                                  \\
              z_2 
        \end{array}  \right)
        = R(\beta)
               \left( \begin{array}{c}
                       z^0                           \\
                       A^0 
                        \end{array}  \right) ,      \nn  
\end{eqnarray}
where $R(\theta)$ is the usual rotation matrix of angle $\theta$.
After diagonalization with setting $\tan \be = v_2 / v_1$, 
the two mass-eigenstates $w^{\pm}$ and $z$ become 
the Nambu-Goldstone bosons which are to be absorbed 
into the longitudinal part of the gauge bosons $W^{\pm}$ and $Z$ respectively. 
The other mass-eigenstates $h^0$, $H^0$, $H^{\pm}$, and $A^0$ become to 
represent five massive Higgs bosons, that is, two CP-even neutral, 
charged and CP-odd neutral ones, respectively.   
Another mixing angle $\al$ is chosen in order that $h^0$ is lighter 
than $H^0$.   
The relation between the coupling constants and masses are 
  \begin{eqnarray}
  \eta_1 
&=& \fr{1}{2 v^2 \cs^2 \be} 
 (\cs^2 \al \;\mH + \sn^2 \al \;\mh - \tan \be \;\mu_3^2), \label{qu1}\\
  \eta_2 
&=& \fr{1}{2 v^2 \sn^2 \be} 
 (\sn^2 \al \;\mH + \cs^2 \al \;\mh - \cot \be \;\mu_3^2), \label{qu2} \\
  \eta_3 
&=& \fr{\sn 2\al}{v^2 \sn 2\be} (\mH - \mh) + \fr{2 \mg}{v^2}
      - \fr{2}{v^2 \sin 2 \be} \;\mu_3^2 ,\label{qu3} \\
  \eta_4 
&=& - \fr{2 \mg}{v^2} + \fr{4}{v^2 \sin 2 \be}\; \mu_3^2, \label{qu4} \\
  \eta_5 
&=& \fr{2}{v^2} (\ma - \mg).
\label{mass}
\end{eqnarray}
Note that since only the $\eta_5$-term explicitly breaks 
the custodial $SU(2)_V$ symmetry in the Higgs sector, Eq \eq{mass} 
implies that the mass splitting between $H^{\pm}$ and $A^0$ 
measures the $SU(2)_V$ breaking. 
The eight independent parameters 
($\mu_1, \mu_2, \mu_3, \eta_1, \sim, \eta_5$) in \eq{int}  
are thus replaced into   
the four mass parameters $m_{h^0}, m_{H^0}, m_{H^\pm}$ and $m_{A^0}$, 
the two mixing angles $\al$ and $\be$, the vacuum expectation value $v$ 
and the soft-breaking parameter $\mu_3$.

Next, we discuss the non-decoupling effects of the Higgs bosons 
in this model. 
In case of the fermion sector, the Yukawa coupling constants 
are naively proportional to fermion masses. 
This implies that the masses of fermions  
are enhanced only by the growing Yukawa coupling constants.  
Then the decoupling theorem does not work and  
the non-decoupling effects of fermion masses appear.   
In case of the SM Higgs sector, the similar effects are expected 
because the quartic self-coupling constant is proportional to 
the Higgs boson mass \cite{ab}.  
On the other hand, in case of the THDM Higgs sector,  
whether the heavy Higgs bosons but the lightest in the loop are 
decoupled or not is a model 
dependent problem. 
For example, the mass of $A^0$ is given from Eqs. \eq{qu4} and \eq{mass} as 
\begin{eqnarray}
  m_{A^0}^2 = \fr{1}{2}(\eta_5 - \eta_4) v^2 
                  + \fr{1}{\sin \beta \cos \be} \;\mu_3^2.
\end{eqnarray}
Naively, $m_{A^0}$ can become large by the growing quartic coupling  
constants or the large soft-breaking parameter $\mu_3^2$. 
If large $m_{A^0}$ is realized by the  
$\mu_3^2$ term with keeping 
the quartic coupling constants to be small, 
the Higgs boson masses then become to be decoupled. 
Note that the MSSM Higgs sector belongs to this type, in which 
all the quartic coupling constants are constrained to ${\cal O}(g^2)$, 
where $g$ is the weak gauge coupling constants,  
so that all the heavy Higgs bosons ($H^0$, $H^\pm$ and $A^0$) 
can grow only due to the large soft-breaking parameters.   
Alternatively, if large $m_A$ is realized by the large quartic coupling 
constants with keeping $\mu_3^2$ to be small, 
the similar situation to the fermion and SM Higgs case occurs. 
Thus non-decoupling contributions of Higgs boson masses are expected 
in these cases \cite{nondec}.   
We note that the non-decoupling Higgs theories often receive 
strong constraints on the Higgs boson masses 
by the perturbative unitarity \cite{uni}.

The situation that Higgs boson masses mainly come from the self-coupling 
constant is, by itself, also seen in the case of SM.
The non-decoupling effects of the Higgs boson 
are then induced in, for example, the oblique corrections. 
However, in the SM case, the Higgs sector with the one Higss doublet 
is custodial $SU(2)_V$ symmetric. 
The leading power-like contributions of  
the Higgs bosons are canceled due to this symmetry and at most  
the sub-leading logarithmic contribution ($\sim \log m_H$) appears. 
This phenomenon in the oblique\\ 
\begin{center}
\begin{picture}(300,120)(0,0)
\SetWidth{1}
\GCirc(110,60){40}{0.5}
\DashLine(10,60)(70,60){10} 
\Photon(142,84)(190,120){4}{4}
\Photon(142,36)(190,0){4}{4}
\LongArrow(162,84)(184,100)
\LongArrow(162,36)(184,20)
\LongArrow(30,70)(50,70)
\Text(190,90)[l]{\Large $p_W$}
\Text(190,30)[l]{\Large $p_Z$}
\Text(35,80)[l]{\Large $p$}
\Text(10,40)[l]{\Large $H^+$}
\Text(205,120)[l]{\Large $W^+_\mu$}
\Text(205,0)[l]{\Large $Z^0_\nu$}
\Text(240,60)[l]{\LARGE $= i g m_W V_{\mu\nu}$}
\end{picture}  
\vspace{1cm}\\
{\large Fig 1.}
\end{center}
\vspace{5mm}
corrections has been known as the screening theorem \cite{sc}.
In THDM, there can be leading power-like contributions 
of Higgs boson masses in the case without the custodial symmetry.     
In this paper, we study the possibility that 
the enhancement of the loop induced $H^\pm W^\mp Z^0$ vertex 
may occur by the similar mechanism.


\section{Non-decoupling Effects on Loop Induced $H^\pm W^\mp Z^0$ vertex}

The $H^\pm W^\mp Z^0$ vertex is defined as $i g m_W V_{\mu\nu}$ 
(See Fig 1.), 
where $V_{\mu\nu}$ is expressed by \cite{mepo}
\begin{eqnarray}
  V_{\mu\nu} = F \gmn + \fr{G}{\mw} {p_Z}_{\mu} {p_W}_{\nu} + 
              \fr{H}{\mw} \epsilon_{\mu\nu\rho\sigma}p_Z^{\rho}p_W^{\sigma}, 
\label{ff}
\end{eqnarray}
where $p_Z$ and $p_W$ are momenta of $Z$ and $W$ bosons, respectively.
All the external lines are assumed to be on mass-shell. 
We have $\partial_\mu W^\mu = 0$ and $\partial_\mu Z^\mu = 0$ then.

First of all, we observe the absence of 
the $H^\pm W^\mp Z^0$ coupling at tree level in THDM.
The tree-level coupling is considered to be generated 
in the kinetic part of the Higgs sector,
\begin{eqnarray}
  {\cal L}_{\rm THDM}^{kin} = \sum_{i =1}^2
   \lt(D_\mu \Phi_i\rt)^\dagger D^\mu \Phi_i, \label{kin1}
\end{eqnarray}
where $D_\mu$ is the covariant derivative for $SU(2)_L \otimes U(1)_Y$.  
In the Georgi basis \cite{georgi}, 
which is obtained from $\Phi_i$ rotating by $\beta$, 
Eq \eq{kin1} becomes 
\begin{eqnarray}
  {\cal L}_{\rm THDM}^{kin} = 
   \lt(D_\mu \Phi\rt)^\dagger D^\mu \Phi +  
   \lt(D_\mu \Psi\rt)^\dagger D^\mu \Psi, \label{kin2}
\end{eqnarray}
where 
\begin{eqnarray}
   \Phi = \left( \begin{array}{c}
                w^+                                             \\
                 \frac{1}{\sqrt{2}} ( \phi^0 + v + i z^0 )
                   \end{array} 
                   \right), 
   \Psi = \left( \begin{array}{c}
                H^+                                              \\
                 \frac{1}{\sqrt{2}} ( \psi^0  + i A^0 )
                   \end{array} 
                   \right), \label{gb}
\end{eqnarray}
and 
\begin{eqnarray}
 \phi^0 &=& \cs \ab H^0 - \sn \ab h^0,\\       
 \psi^0 &=& \sn \ab H^0 + \cs \ab h^0.
\end{eqnarray}
Since $\Psi$ does not have any vacuum expectation value and 
there is no mixing between $\Phi$ and $\Psi$ in Eq \eq{kin2}, 
we can understand the absence of $H^\pm W^\mp Z^0$ at the tree level. 
The $H^\pm W^\mp Z^0$ vertex can induced only by the loop level 
where the mixing between $\Phi$ and $\Psi$ is induced   
through effects from the other sectors.

Second, we discuss the non-decoupling effects of heavy particles on 
the vertex.  The effective Lagrangian is 
\begin{eqnarray}
  {\cal L}_{\rm eff} 
  &=& f_{H^+ W^- Z^0} H^+ W^-_\mu Z^\mu + {\rm h.c.} \nn\\
  &+& g_{H^+ W^- Z^0} H^+ F^{\mu\nu}_Z F_{\mu\nu}^W + {\rm h.c.}\nn \\
  &+& h_{H^+ W^- Z^0} \;\;  i \epsilon_{\mu\nu\rho\sigma} 
      H^+ F^{\mu\nu}_Z F^{\rho\sigma}_W  + {\rm h.c.}. \label{ef}
\end{eqnarray}
Since the coefficient $f_{H^+ W^- Z^0}$ is mass-dimension one,  
the contribution of the heavy particles with the masses $M_i$ 
take the form at one loop level like 
\begin{eqnarray}
   f_{H^+ W^- Z^0} 
       \sim g \times \fr{g}{\cos \theta_W} \times 
                    \fr{M_i^2}{v} \times f(M_i)
       \sim \fr{m_W m_Z}{v^3} \times M^2_i f(M_i),      
\end{eqnarray}
where $f(M_i)$ is a dimensionless function of $M_i$'s.  
Hence  the leading contributions of heavy masses to $f_{H^+ W^- Z^0}$ 
are expected to be quadratic ones. 
As for the dimension $-1$ coefficients 
$g_{H^+ W^- Z^0}$ and $h_{H^+ W^- Z^0}$, they are expected 
to take the forms at one loop level like   
\begin{eqnarray}
  g_{H^+ W^- Z^0}, \; h_{H^+ W^- Z^0} \sim \fr{m_W m_Z}{v^3} \times f'(M_i),
\end{eqnarray}
where $f'(M_i)$ is a dimensionless function of $M_i$'s. 
Namely they have the Higgs mass contributions like at most $\sim \log M_i$.
Therefore from the naive power counting, we expect that there 
may be non-decoupling effects (quadratic and logarithmic mass contributions) 
of the heavy Higgs bosons as well as heavy fermions 
at the one loop induced $H^\pm W^\mp Z^0$ vertex 
\footnote
{Since there is no correspondence to $f_{H^+W^-Z^0}$ term 
in the effective Lagrangian for the $H^\pm W^\mp \gamma$ vertex 
because of the gauge invariant condition $p_\gamma^\mu V_{\mu\nu} = 0$, 
the quadratic mass effects disappear in the one loop induced 
$H^\pm W^\mp \gamma$ vertex \cite{capd}.
Thus the large enhancement of the vertex cannot be expected 
in this vertex and this fact is the reason what we consider only 
the $H^\pm W^\mp Z^0$ vertex.}.

Third, 
we show that the the non-decoupling effects on the vertex is strongly 
constrained if there is the custodial $SU(2)_V$ symmetry.   
In our model, the effective Lagrangian \eq{ef} comes from the operators, 
\begin{eqnarray}
{\rm tr}\lt[ \tau_3 (D_\mu {\cal M})^\dagger (D^\mu {\cal N}) \rt], \;\;
{\rm tr}\lt[ \tau_3 {\cal M}^\dagger {\cal N} 
                             F_Z^{\mu\nu} F_{\mu\nu}^W \rt] \; 
{\rm and}\;\;\;i \epsilon_{\mu\nu\rho\sigma}
     {\rm tr}\lt[ \tau_3 {\cal M}^\dagger {\cal N}
                     F_Z^{\mu\nu} F^{\rho\sigma}_W \rt], \label{op}
\end{eqnarray}
where $2 \times 2$ matrices ${\cal M}$ and ${\cal N}$ are defined by 
using the doublets $\Phi$ and $\Psi$ in \eq{gb} as
${\cal M}= (i\tau_2 \Phi^\ast, \Phi)$ and 
${\cal N} = (i\tau_2 \Psi^\ast,\Psi)$. 
Since it can be seen clearly that 
all the operators \eq{op} are un-invariant under $SU(2)_R$ and thus 
$SU(2)_V$, we understand that the vertex appears only 
in the case without the custodial $SU(2)_V$ symmetry.

In our model, there are three independent sources for the explicit 
$SU(2)_V$ breaking according to the gauge, fermion and Higgs sectors. 
They are separately measured by the three mass-splittings 
$\mw - \mz$, $m_t^2 - m_b^2$ and $m_A^2 - m_{H^\pm}^2$, respectively.
Since the breakdown in the gauge as well as the fermion sectors is 
already known experimentally,  
the quadratic mass contributions of the gauge bosons and fermions 
appear in the vertex.   
In MSSM, 
only the effects of the heavy fermions (especially the top-quark) 
become important because the explicit breaking of $SU(2)_V$ 
in both the gauge and Higgs sectors are small.
However, in the present situation that the top-quarks has been already 
discovered with the mass of $\sim 175$ GeV \cite{top} 
and the forth generation of quarks has almost been  
excluded by the S-parameter constraints \cite{hmh,4gen}, 
it seems to be difficult to 
obtain a substantial enhancement of the vertex only by the fermion effects. 
Hence the only hope to the novel enhancement of the vertex is 
due to the non-decoupling Higgs sector 
with enough large mass splitting $m_{A^0} - m_{H^\pm}$.
 
Finally we note that, as well known, the experimental value of 
$\rho$ parameter gives a strong constraint for the breaking of 
the custodial symmetry.    
Even in THDM, the large parameter region have been already 
excluded by the data 
\cite{hmh,gr}. 
We, however, stress that the $\rho$ parameter constraint does not always 
forbid large $SU(2)_V$ breaking in the THDM Higgs sector. 
For one of the examples, 
if we consider the case with $\al -\be \sim \pi/2$ and 
$\mg \sim \mH$, the large mass splitting between $A^0$ and $H^\pm$ 
is possible with keeping $\Delta \rho = \rho - 1 \sim 0$.

Thus it becomes important and very interesting to study 
the non-decoupling effects of the Higgs bosons on the vertex. 
It is possible 
that $m_A^2 - m_{H^\pm}^2$ is large enough to give substantial 
non-decoupling effects such as $\sim M_i^2$  on the vertex 
with keeping $\Delta \rho \sim 0$.

\section{Analysis of $H^+ \rightarrow W^+Z^0$}

\hspace*{18pt}
In Sec 3, we have qualitatively discussed the possibility 
that the loop-induced 
$H^\pm W^\mp Z^0$ vertex can be significantly enhanced by the Higgs 
non-decoupling effects if there is large explicit breaking of 
the custodial $SU(2)_V$ symmetry in the Higgs sector. 
To verify this observation quantitatively, 
we proceed to the analysis of the decay process $H^+ \rightarrow W^+ Z^0$.  
We here present the one-loop calculation of the decay width 
and evaluate the branching ratio of this decay mode. 
The decay width is given in terms of the form factors $F$, $G$ and $H$ 
(see Eq \eq{ff}) as \cite{mepo}
\begin{eqnarray}
  \Gamma (H^+ \rightarrow W^+ Z^0)
    = m_{H^\pm}
  \fr{\lambda^{1/2}(1,w,z)}{16 \pi} 
  \lt( |{\cal M}_{LL}|^2 + |{\cal M}_{TT}|^2 \rt),  \label{width}
\end{eqnarray}
where  $w = (m_W/m_{H^\pm})^2$, $z = (m_Z/m_{H^\pm})^2$,
 $\lambda (a,b,c) = (a - b - c)^2 - 4 b c$.
The amplitudes ${\cal M}_{LL}$ and ${\cal M}_{TT}$ are 
the contributions of each modes of 
the longitudinally and transversely polarized final 
gauge bosons.   We have their explicit expressions,        
\begin{eqnarray}
  |{\cal M}_{LL}|^2 &=& \fr{g^2}{4 z}
    \lt| (1 - w - z) F + \fr{\lambda (1,w,z)}{2w} G \rt|^2 \label{ll}\\
  |{\cal M}_{TT}|^2 &=& g^2 
     \lt\{ 2w |F|^2 + \fr{\lambda (1,w,z)}{2w} |H|^2 \rt\}. \label{tt}  
\end{eqnarray}
The contributions of the diagrams with 
a boson (Higgs, Nambu-Goldstone or gauge bosons) loop  
are calculated by employing the t'Hooft-Feynman gauge here.  
The boson-loop diagrams are shown in Figs 2(a), (b) and (c).
The explicit expressions of these boson-loop contributions
to the quantities $F$ and $G$ are given in Appendix 1.
All the boson-loop diagrams do not contribute to the quantity $H$ at all 
because the boson sectors of the theory \\
\begin{center}
\begin{picture}(400,480)(0,50)
\SetWidth{1}
\Vertex(40,450){1} \DashLine(10,450)(40,450){5}  
\Vertex(80,480){1}\Photon(80,480)(110,480){3}{3}
\Vertex(80,420){1}\Photon(80,420)(110,420){3}{3}
\Text(0,440)[l]{\large $H^+$}
\Text(115,480)[l]{\large $W^+_\mu$}
\Text(115,420)[l]{\large $Z^0_\nu$}
\DashLine(40,450)(80,480){5}  
\DashLine(40,450)(80,420){5}  
\DashLine(80,480)(80,420){5}        
\Text(50,475)[l]{\large $H^\pm$}
\Text(50,425)[c]{\large $h^0,H^0$}
\Text(85,450)[l]{\large $A^0$}     
\Vertex(170,450){1} \DashLine(140,450)(170,450){5}  
\Vertex(210,480){1}\Photon(210,480)(240,480){3}{3}
\Vertex(210,420){1}\Photon(210,420)(240,420){3}{3}
\DashLine(170,450)(210,480){5} 
\DashLine(170,450)(210,420){5} 
\DashLine(210,480)(210,420){5}
\Text(180,475)[c]{\large $h^0,H^0$}
\Text(180,425)[l]{\large $H^\pm$}
\Text(215,450)[l]{\large $H^\pm$}     
\Vertex(300,450){1} \DashLine(270,450)(300,450){5}  
\Vertex(340,480){1}\Photon(340,480)(370,480){3}{3}
\Vertex(340,420){1}\Photon(340,420)(370,420){3}{3}
\Photon(300,450)(340,480){3}{5} 
\DashLine(300,450)(340,420){5} 
\DashLine(340,480)(340,420){5}
\Text(310,480)[l]{\large $W^\pm$}
\Text(310,425)[l]{\large $A^0$}
\Text(345,450)[l]{\large $h^0,H^0$}     
\Vertex(40,330){1} \DashLine(10,330)(40,330){5}  
\Vertex(80,360){1}\Photon(80,360)(110,360){3}{3}
\Vertex(80,300){1}\Photon(80,300)(110,300){3}{3}
\DashLine(40,330)(80,360){5} 
\Photon(40,330)(80,300){3}{5} 
\DashLine(80,360)(80,300){5}
\Text(50,355)[l]{\large $H^\pm$}
\Text(50,300)[l]{\large $Z^0$}
\Text(85,330)[l]{\large $h^0,H^0$}     
\Vertex(170,330){1} \DashLine(140,330)(170,330){5}  
\Vertex(210,360){1}\Photon(210,360)(240,360){3}{3}
\Vertex(210,300){1}\Photon(210,300)(240,300){3}{3}
\DashLine(170,330)(210,360){5} 
\Photon(170,330)(210,300){3}{5} 
\Photon(210,300)(210,360){3}{5}
\Text(180,360)[c]{\large $h^0,H^0$}
\Text(180,300)[l]{\large $W^\pm$}
\Text(215,330)[l]{\large $W^\pm$}     
\Vertex(300,330){1} \DashLine(270,330)(300,330){5}  
\Vertex(340,360){1}\Photon(340,360)(370,360){3}{3}
\Vertex(340,300){1}\Photon(340,300)(370,300){3}{3}
\Photon(300,330)(340,360){3}{5} 
\DashLine(300,330)(340,300){5} 
\Photon(340,360)(340,300){3}{5}
\Text(310,365)[l]{\large $W^\pm$}
\Text(310,305)[c]{\large $h^0,H^0$}
\Text(345,330)[l]{\large $Z^0$}     
\Vertex(40,210){1} \DashLine(10,210)(40,210){5}  
\Vertex(80,240){1}\Photon(80,240)(110,240){3}{3}
\Vertex(80,180){1}\Photon(80,180)(110,180){3}{3}
\DashLine(40,210)(80,240){5} 
\DashLine(40,210)(80,180){5} 
\DashLine(80,240)(80,180){5}
\Text(50,235)[l]{\large $w^\pm$}
\Text(50,185)[l]{\large $A^0$}
\Text(85,210)[l]{\large $h^0,H^0$}    
\Vertex(170,210){1} \DashLine(140,210)(170,210){5}  
\Vertex(210,240){1}\Photon(210,240)(240,240){3}{3}
\Vertex(210,180){1}\Photon(210,180)(240,180){3}{3}
\DashLine(170,210)(210,240){5} 
\DashLine(170,210)(210,180){5} 
\DashLine(210,240)(210,180){5}
\Text(180,235)[l]{\large $w^\pm$}
\Text(180,180)[c]{\large $h^0,H^0$}
\Text(215,210)[l]{\large $z^0$}     
\Vertex(300,210){1} \DashLine(270,210)(300,210){5}  
\Vertex(340,240){1}\Photon(340,240)(370,240){3}{3}
\Vertex(340,180){1}\Photon(340,180)(370,180){3}{3}
\DashLine(300,210)(340,240){5} 
\DashLine(300,210)(340,180){5} 
\DashLine(340,240)(340,180){5}
\Text(310,240)[c]{\large $h^0,H^0$}
\Text(310,180)[l]{\large $w^\pm$}
\Text(345,210)[l]{\large $w^\pm$}
\Vertex(40,90){1} \DashLine(10,90)(40,90){5}  
\Vertex(80,120){1}\Photon(80,120)(110,120){3}{3}
\Vertex(80,60){1}\Photon(80,60)(110,60){3}{3}
\DashLine(40,90)(80,120){5} 
\DashLine(40,90)(80,60){5} 
\Photon(80,120)(80,60){3}{5}
\Text(50,115)[l]{\large $w^\pm$}
\Text(50,65)[c]{\large $h^0,H^0$}
\Text(85,90)[l]{\large $Z^0$}    
\Vertex(170,90){1} \DashLine(140,90)(170,90){5}  
\Vertex(210,120){1}\Photon(210,120)(240,120){3}{3}
\Vertex(210,60){1}\Photon(210,60)(240,60){3}{3}
\DashLine(170,90)(210,120){5} 
\DashLine(170,90)(210,60){5} 
\Photon(210,120)(210,60){3}{5}
\Text(180,115)[c]{\large $h^0,H^0$}
\Text(180,60)[l]{\large $w^\pm$}
\Text(215,90)[l]{\large $W^\pm$}     
\Vertex(300,90){1} \DashLine(270,90)(300,90){5}  
\Vertex(340,120){1}\Photon(340,120)(370,120){3}{3}
\Vertex(340,60){1}\Photon(340,60)(370,60){3}{3}
\DashLine(300,90)(340,120){5} 
\Photon(300,90)(340,60){3}{5} 
\DashLine(340,120)(340,60){5}
\Text(310,120)[c]{\large $h^0,H^0$}
\Text(310,60)[l]{\large $W^\pm$}
\Text(345,90)[l]{\large $w^\pm$}
\end{picture}  
\vspace{5mm}\\
{\Large Fig 2-a}
\end{center}
\vspace*{5mm}

\noindent
has the parity ($P$) symmetry. 
The fermion-loop diagrams are shown in Fig 2(d). 
Since all the diagrams with a fermion-loop themselves
construct a gauge invariant subset, we are free to use
different gauge choices for boson- and fermion-loops.
It is clear that the unitary gauge is the most convenient choice
for the contributions of the fermion-loop diagrams. 
The explicit calculations of the fermion-loop contribution 
are given in Ref \cite{capd}.  We have \\
\begin{center}
\begin{picture}(400,350)(0,0)
\SetWidth{1}
\Text(0,280)[l]{\large $H^+$}
\Text(115,315)[l]{\large $W^+_\mu$}
\Text(115,265)[l]{\large $Z^0_\nu$}
\Vertex(40,290){1} 
\DashLine(10,290)(40,290){5}  
\DashCArc(60,290)(20,0,180){5}
\DashCArc(60,290)(20,180,360){5}
\Vertex(80,290){1}
\Photon(80,290)(110,315){3}{4}
\Photon(80,290)(110,265){3}{4}
\Text(60,320)[c]{\large $H^\pm, w^\pm$}
\Text(60,260)[c]{\large $h^0,H^0$}
\DashLine(155,280)(185,280){5}  \Vertex(185,280){1}
\Photon(215,250)(185,280){2}{5}
\DashCArc(199.14,294.14)(20,225,45){5}
\PhotonArc(199.14,294.14)(20,45,225){2}{7}
\Vertex(213.28,308.28){1}
\Photon(213.28,308.28)(233.28,328.28){2}{3}
\Text(175,315)[c]{\large $W^\pm$}
\Text(220,280)[l]{\large $h^0,H^0$}
\DashLine(280,300)(310,300){5}  \Vertex(310,300){1}
\Photon(340,330)(310,300){2}{5}
\DashCArc(324.14,285.86)(20,315,135){5}
\PhotonArc(324.14,285.86)(20,135,315){2}{7}
\Vertex(338.28,271.72){1}
\Photon(338.28,271.72)(358.28,251.72){2}{3}
\Text(300,265)[c]{\large $Z^0$}
\Text(345,300)[l]{\large $h^0,H^0$}

\Vertex(40,170){1} 
\DashLine(10,170)(40,170){5}  
\PhotonArc(60,170)(20,0,180){2}{7}
\DashCArc(60,170)(20,180,360){5}
\Vertex(80,170){1}
\Photon(80,170)(120,170){2}{4}  
\Vertex(120,170){1}
\Photon(120,170)(150,195){2}{4}
\Photon(120,170)(150,145){2}{4}
\Text(60,200)[c]{\large $W^\pm$}
\Text(60,140)[c]{\large $h^0,H^0$}
\Text(100,160)[c]{\large $W^\pm$}
\Vertex(240,170){1} 
\DashLine(210,170)(240,170){5}  
\DashCArc(260,170)(20,0,180){5}
\DashCArc(260,170)(20,180,360){5}
\Vertex(280,170){1}
\Photon(280,170)(320,170){2}{4}  
\Vertex(320,170){1}
\Photon(320,170)(350,195){2}{4}
\Photon(320,170)(350,145){2}{4}
\Text(260,200)[c]{\large $H^\pm,w^\pm$}
\Text(260,140)[c]{\large $h^0,H^0$}
\Text(300,160)[c]{\large $W^\pm$}

\Vertex(40,55){1} 
\DashLine(10,55)(40,55){5}  
\PhotonArc(60,55)(20,0,180){2}{7}
\DashCArc(60,55)(20,180,360){5}
\Vertex(80,55){1}
\DashLine(80,55)(120,55){5}  
\Vertex(120,55){1}
\Photon(120,55)(150,80){2}{4}
\Photon(120,55)(150,30){2}{4}
\Text(60,85)[c]{\large $W^\pm$}
\Text(60,25)[c]{\large $h^0,H^0$}
\Text(100,45)[c]{\large $w^\pm$}
\Vertex(240,55){1} 
\DashLine(210,55)(240,55){5}  
\DashCArc(260,55)(20,0,180){5}
\DashCArc(260,55)(20,180,360){5}
\Vertex(280,55){1}
\DashLine(280,55)(320,55){5}  
\Vertex(320,55){1}
\Photon(320,55)(350,80){2}{4}
\Photon(320,55)(350,30){2}{4}
\Text(260,85)[c]{\large $H^\pm,w^\pm$}
\Text(260,25)[c]{\large $h^0,H^0$}
\Text(300,45)[c]{\large $w^\pm$}
\end{picture}  \\
{\Large Fig 2-b}
\end{center}
\vspace*{5mm}
just checked their results.

\subsection{Heavy mass limit}

\hspace*{18pt}
We consider to extract the contribution of the masses of heavy Higgs bosons.
In this subsection, we assume that both the neutral ($h^0$, $H^0$) 
and CP-odd ($A^0$) Higgs bosons are heavier than the charged one. 
The soft-breaking parameter $\mu_3^2$ is also set into zero for a while  
in order to obtain the non-decoupling effects of the Higgs bosons maximally.

A naive counting by using Eqs \eq{ll} and \eq{tt} 
shows that the ratio $|{\cal M}_{TT}/{\cal M}_{LL}|^2$ 
behaves like $\sim 8 \cdot \mw/\mg \cdot \mz/\mg$.    
If $\mg/\mw$ is large, the contribution of ${\cal M}_{LL}$ 
becomes dominant. 
The bosonic loop contributions to ${\cal M}_{LL}$  
can rewritten by factorizing the mixing\\
\begin{center}
\begin{picture}(400,220)(0,0)
\SetWidth{1}

\DashLine(30,170)(90,170){5}  \Vertex(90,170){1}
\Text(30,160)[c]{\large $H^+$}
\Photon(120,200)(90,170){2}{5}
\Text(130,210)[c]{\large $W^+_{\mu}$}
\Photon(90,170)(120,140){2}{5}
\Text(130,130)[c]{\large $Z^0_{\nu}$}
\DashLine(90,170)(72.68,200){5} \Vertex(71.98,199){2}
\GCirc(65.78,212){13.866}{0.5}
\Text(55,185)[c]{\large $h^0,H^0$}

\Vertex(250,170){2} 
\DashCArc(250,185)(15,0,180){5}
\DashCArc(250,185)(15,180,360){5}
\Text(250,215)[c]{\large $H^\pm$, $h^0$, $H^0$, $A^0$, $w^\pm$, $z^0$}

\DashLine(210,170)(290,170){5}  
\Vertex(290,170){1}
\Photon(290,170)(320,195){2}{4}
\Photon(290,170)(320,145){2}{4}
\Text(275,160)[c]{\large $w^\pm$}
\Vertex(80,50){1} 
\DashLine(80,50)(80,80){5}
\Vertex(80,79){2}
\GCirc(80,95){15}{0.5}
\DashLine(40,50)(80,50){5}
\Photon(80,50)(120,50){2}{4}  
\Vertex(120,50){1}
\Photon(120,50)(150,75){2}{4}
\Photon(120,50)(150,25){2}{4}
\Text(100,35)[c]{\large $W^\pm$}
\Text(85,70)[l]{\large $h^0,H^0$}
\Vertex(250,50){1} 
\DashLine(250,50)(250,80){5}
\Vertex(250,79){2}
\GCirc(250,95){15}{0.5}
\DashLine(210,50)(250,50){5}
\DashLine(250,50)(290,50){5}  
\Vertex(290,50){1}
\Photon(290,50)(320,75){2}{4}
\Photon(290,50)(320,25){2}{4}
\Text(275,40)[c]{\large $w^\pm$}
\Text(255,70)[l]{\large $h^0,H^0$}
\end{picture}  
\begin{picture}(400,60)(0,5)
\SetWidth{1}
\Text(10,50)[c]{\Large where} 
\Vertex(50,9){3} 
\GCirc(50,25){15}{0.5}

\Text(105,20)[c]{\Large $=$} 
\Vertex(160,9){3}
\DashCArc(160,25)(15,0,180){5}
\DashCArc(160,25)(15,180,360){5}
\Text(190,40)[l]{\large $H^\pm$, $h^0$, $H^0$, $A^0$, $w^\pm$, $z^0$, 
                $W^\pm$, $Z^0$}
\end{picture}  
\vspace{5mm}\\
{\Large Fig 2-c}
\end{center}
\vspace{5mm}
 angle dependence as 
\begin{eqnarray}
  {\cal M}_{LL} = 
  J(\al,\be)  {\cal M}_{LL}^J(M_i) 
+ K(\al,\be)  {\cal M}_{LL}^K(M_i)
+ L(\al,\be)  {\cal M}_{LL}^L(M_i),  
  \label{fac}
\end{eqnarray}
where $M_i$ represent the masses of $h^0$, $H^0$ and $A^0$, and
\begin{eqnarray}
J(\al,\be) &=& \snab \csab,                          \\
K(\al,\be) &=& \sin^2 \al \ctb - \cos^2 \al \tnb,    \\
L(\al,\be) &=& \cos^2 \al \ctb - \sin^2 \al \tnb. 
\end{eqnarray}
The leading (quadratic) mass effects of heavier Higgs bosons on 
${\cal M}_{LL}^J(M_i)$, 
${\cal M}_{LL}^K(M_i)$ and ${\cal M}_{LL}^L(M_i)$ 
are then extracted from the full expression in Appendix 1 as  
\begin{eqnarray}
   {\cal M}_{LL}^J(M_i) &\sim& 0,   \label{fa1} \\  
   {\cal M}_{LL}^K(M_i) &\sim&  
       \fr{m_{H^\pm}}{2 (4 \pi)^2 v^3} \times  
        \fr{\mH \ma}{\mH - \ma} \ln \fr{\mH}{\ma}, \label{fa2} \\  
   {\cal M}_{LL}^K(M_i) &\sim& 
       \fr{m_{H^\pm}}{2 (4 \pi)^2 v^3} \times  
        \fr{\mh \ma}{\mh - \ma} \ln \fr{\mh}{\ma}. \label{fa3}  
\end{eqnarray}

\begin{center}
\begin{picture}(400,70)(0,0)
\SetWidth{1}
\Text(0,25)[l]{\large $H^+$}
\Text(120,60)[l]{\large $W^+_\mu$}
\Text(120,10)[l]{\large $Z^0_\nu$}
\Vertex(20,35){1} 
\DashLine(0,35)(20,35){5}  
\CArc(40,35)(20,0,180)
\CArc(40,35)(20,180,360)
\Vertex(60,35){1}
\Photon(60,35)(90,35){2}{4}  
\Vertex(90,35){1}
\Photon(90,35)(115,60){2}{4}
\Photon(90,35)(115,10){2}{4}
\Text(40,65)[c]{\large $t$}
\Text(40,5)[c]{\large $b$}
\Text(80,25)[c]{\large $W^\pm$}
\Vertex(180,35){1} \DashLine(150,35)(180,35){5}  
\Vertex(220,65){1} \Photon(220,65)(250,65){3}{3}
\Vertex(220,5){1} \Photon(220,5)(250,5){3}{3}
\Line(180,35)(220,65) 
\Line(180,35)(220,5) 
\Line(220,65)(220,5)
\Text(190,60)[l]{\large $t$}
\Text(190,10)[l]{\large $b$}
\Text(225,35)[l]{\large $b$}     
\Vertex(315,35){1} \DashLine(285,35)(315,35){5}  
\Vertex(355,65){1}\Photon(355,65)(385,65){3}{3}
\Vertex(355,5){1}\Photon(355,5)(385,5){3}{3}
\Line(315,35)(355,65)
\Line(315,35)(355,5) 
\Line(355,65)(355,5)
\Text(325,60)[l]{\large $b$}
\Text(325,10)[l]{\large $t$}
\Text(360,35)[l]{\large $t$}     
\end{picture}  
\vspace{5mm}\\
{\Large Fig 2-d}
\end{center}

\noindent
Eq \eq{fac} shows that the vertex can be strongly enhanced 
by the Higgs mass effects if $\tnb$ or $\ctb$ is large enough. 
\footnote{
 In MSSM, the soft-breaking term $\mu_3^2$ 
 cannot be neglected and the mixing angles are not independent of each other. 
 In the large mass limit, we have 
 $m_{A^0}^2 \sim m_{H^\pm}^2 \sim m_{H^0}^2 \sim 2 \mu_3^2/\sin 2 \beta$      
 and $\al = \be - \pi/2$. 
 The enhancement mentioned above can no longer appear in this case 
 because of the cancellation due to the soft breaking parameter. 
 In general THDM, this cancellation do not have to take place as 
 in the cases above.}

The results in Eqs \eq{fa1} $\sim$ \eq{fa3} are also reproduced from 
the much simpler calculation of $H^+ \rightarrow w^+ z^0$ 
by virtue of the equivalence theorem \cite{et0,et}.    
The calculation in this way is performed by the Landau gauge \cite{kt} and 
the diagrams are shown in Fig 3.   
We show the explicit form of the amplitude 
calculated in this way in Appendix 2. 
To take the heavy mass limit in this amplitude leads to 
the completely same results as Eqs \eq{fa1} $\sim$ \eq{fa3}.   
The use of the equivalence theorem is very useful to  
check the results of the full calculation (See Fig 7(a) and (b).).
One more advantage of the use of the equivalence theorem here 
is the fact that we can immediately see the disappearance 
of the amplitudes if the Higgs sector is custodial $SU(2)_V$ symmetric. 
All the diagrams in Fig 3 have a coupling of $H^\pm w^\mp A^0$ 
(shown by the brack dot in each diagram). 
The coupling constant is just $\eta_5$ 
($\propto m_{A^0}^2 - m_{H^\pm}^2 $) which represents 
the explicit breaking of $SU(2)_V$ in the Higgs sector.

\subsection{Numerical Estimation}

\hspace{18pt}
The decay width $\Gamma(H^+ \rightarrow W^+ Z^0)$ is evaluated by using 
Eqs \eq{width}, \eq{ll} and \eq{tt}.
For comparison, the MSSM (with heavy sparticles) case and 
the non-decoupling THDM case are 
shown in Fig 4(a) and 4(b), respectively. 
In MSSM, the heavy Higgs bosons are approximately degenerated 
($m_{H^\pm} - m_{A^0} < 11$ GeV for $m_{H^\pm} > 300$ GeV). 
Hence the Higgs\\
\begin{center}
\begin{picture}(400,305)(0,50)
\SetWidth{1}
\Vertex(40,270){1}\DashLine(10,270)(40,270){5}  
\Text(0,260)[l]{\large $H^+$}
\Vertex(80,300){5}\DashLine(80,300)(110,300){5}
\Vertex(80,240){1}\DashLine(80,240)(110,240){5}
\Text(120,300)[l]{\large $w^+$}
\Text(120,240)[l]{\large $z^0$}
\DashLine(40,270)(80,300){5} 
\DashLine(40,270)(80,240){5} 
\DashLine(80,300)(80,240){5}
\Text(50,295)[l]{\large $H^\pm$}
\Text(50,240)[c]{\large $h^0,H^0$}
\Text(85,270)[l]{\large $A^0$}     
\Vertex(220,270){5}\DashLine(190,270)(220,270){5}  
\Vertex(260,300){1}\DashLine(260,300)(290,300){5}
\Vertex(260,240){1}\DashLine(260,240)(290,240){5}
\DashLine(220,270)(260,300){5} 
\DashLine(220,270)(260,240){5} 
\DashLine(260,300)(260,240){5}
\Text(230,295)[l]{\large $w^\pm$}
\Text(230,240)[l]{\large $A^0$}
\Text(265,270)[l]{\large $h^0,H^0$}     
\Vertex(80,180){5} 
\DashLine(50,180)(80,180){5}  
\DashCArc(100,180)(20,0,180){5}
\DashCArc(100,180)(20,180,360){5}
\Vertex(120,180){1}
\DashLine(120,180)(150,205){5}
\DashLine(120,180)(150,155){5}
\Text(100,210)[c]{\large $w^\pm$}
\Text(100,150)[c]{\large $A^0$}
\Vertex(220,180){1} 
\DashLine(190,180)(220,180){5}  
\DashCArc(240,180)(20,0,180){5}
\DashCArc(240,180)(20,180,360){5}
\Vertex(260,180){5}
\DashLine(260,180)(290,205){5}
\DashLine(260,180)(290,155){5}
\Text(240,210)[c]{\large $H^\pm$}
\Text(240,150)[c]{\large $h^0,H^0$}
\DashLine(15,80)(45,80){5}  \Vertex(45,80){1}
\DashLine(75,50)(45,80){5}
\DashCArc(59.14,94.14)(20,225,45){5}
\DashCArc(59.14,94.14)(20,45,225){5}
\Vertex(73.28,108.28){5}
\DashLine(73.28,108.28)(93.28,128.28){5}
\Text(35,115)[c]{\large $H^\pm$}
\Text(80,80)[l]{\large $A^0$}
\DashLine(150,80)(180,80){5}  \Vertex(180,80){5}
\DashLine(210,50)(180,80){5}
\DashCArc(194.14,94.14)(20,225,45){5}
\DashCArc(194.14,94.14)(20,45,225){5}
\Vertex(208.28,108.28){1}
\DashLine(208.28,108.28)(228.28,128.28){5}
\Text(170,115)[c]{\large $w^\pm$}
\Text(215,80)[l]{\large $h^0,H^0$}
\DashLine(280,100)(310,100){5}  \Vertex(310,100){5}
\DashLine(340,130)(310,100){5}
\DashCArc(324.14,85.86)(20,315,135){5}
\DashCArc(324.14,85.86)(20,135,315){5}
\Vertex(338.28,71.72){1}
\DashLine(338.28,71.72)(358.28,51.72){5}
\Text(295,65)[c]{\large $z^0,A^0$}
\Text(345,100)[l]{\large $h^0,H^0$}
\end{picture}  
\vspace{1cm}\\
{\Large Fig 3}
\end{center}

\noindent
effects are small and heavy 
fermion (top-quark) effects are dominant. 
Since the $H^\pm tb$ coupling consists of 
$\sim m_t \cot \be$ and $m_b \tan \be$, 
the top-quark contributions are rapidly reduced for larger $\tan \be$.  
We can see from Fig 4(a) that $\Gamma(H^+ \rightarrow W^+ Z^0) < 10^{-3}$ 
at $m_{H^\pm} =300$ GeV for $\tan \be > 1$. 
These results for MSSM are quitely consistent with 
the previous ones \cite{mepo,capd}. 
On the other hand, in Fig 2(b), 
in case of THDM with $m_{H^\pm} - m_{A^0} = 200$ GeV, 
the novel enhancement of the width is realized for large $\tan \be$ 
due to the Higgs non-decoupling effects 
(we are setting $\mu_3$ into zero here).
In Fig 5, we show the $\tan \be$ dependence of 
$\Gamma(H^+ \rightarrow W^+ Z^0)$ for various mass splitting 
$\Delta m = |m_{A^0} - m_{H^\pm}|$. 
The enhancement in small $\tan \be$ region ($\tan \beta < 1$) 
is due to the top quark contribution, 
while the width can be considerably enhanced even for large $\tan \be$ 
regions by the Higgs non-decoupling effects if $\Delta m$ become large enough.

We next consider the branching ratio. 
The decay mode is kinematically allowed if $m_{H^\pm} > m_W + m_Z$. 
This is very close to the threshold of $tb$ mode 
with the top-quark mass\\
\begin{minipage}[t]{7cm}  
\epsfxsize=8.5cm
\epsfbox{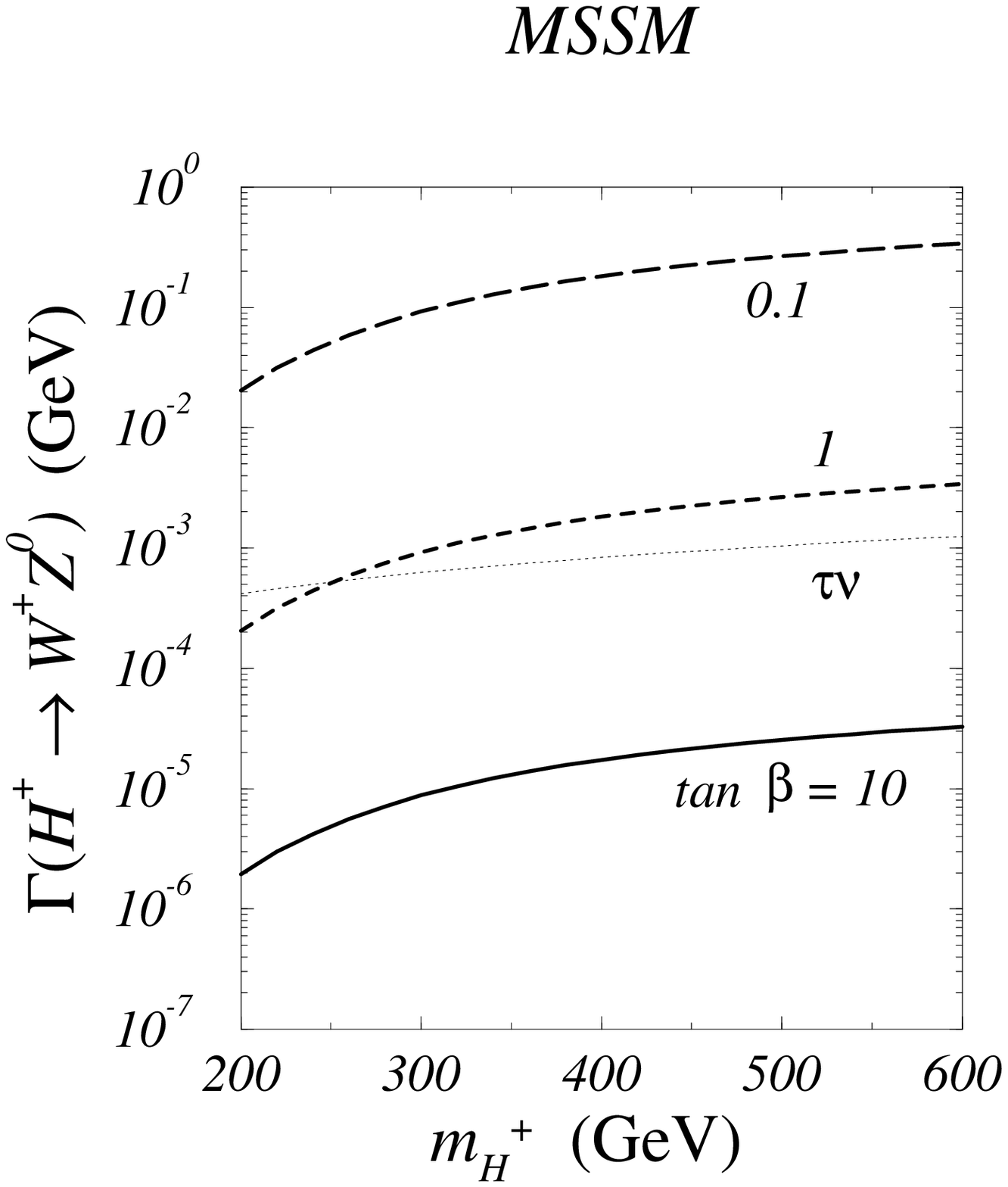}
  {\bf Fig 4(a):} Decay width $\Gamma(H^+\rightarrow W^+Z^0)$
                  for $\tan \be = 0.1, 1, 10$ in MSSM 
                  (All the sparticles are assumed to be very heavy).
                  The decay width of $H^+ \rightarrow \bar{\tau} \nu$
                  for $\tan\be =1$ is also shown for comparison. 
\end{minipage}
\begin{minipage}[t]{7cm}
\epsfxsize=8.5cm
\epsfbox{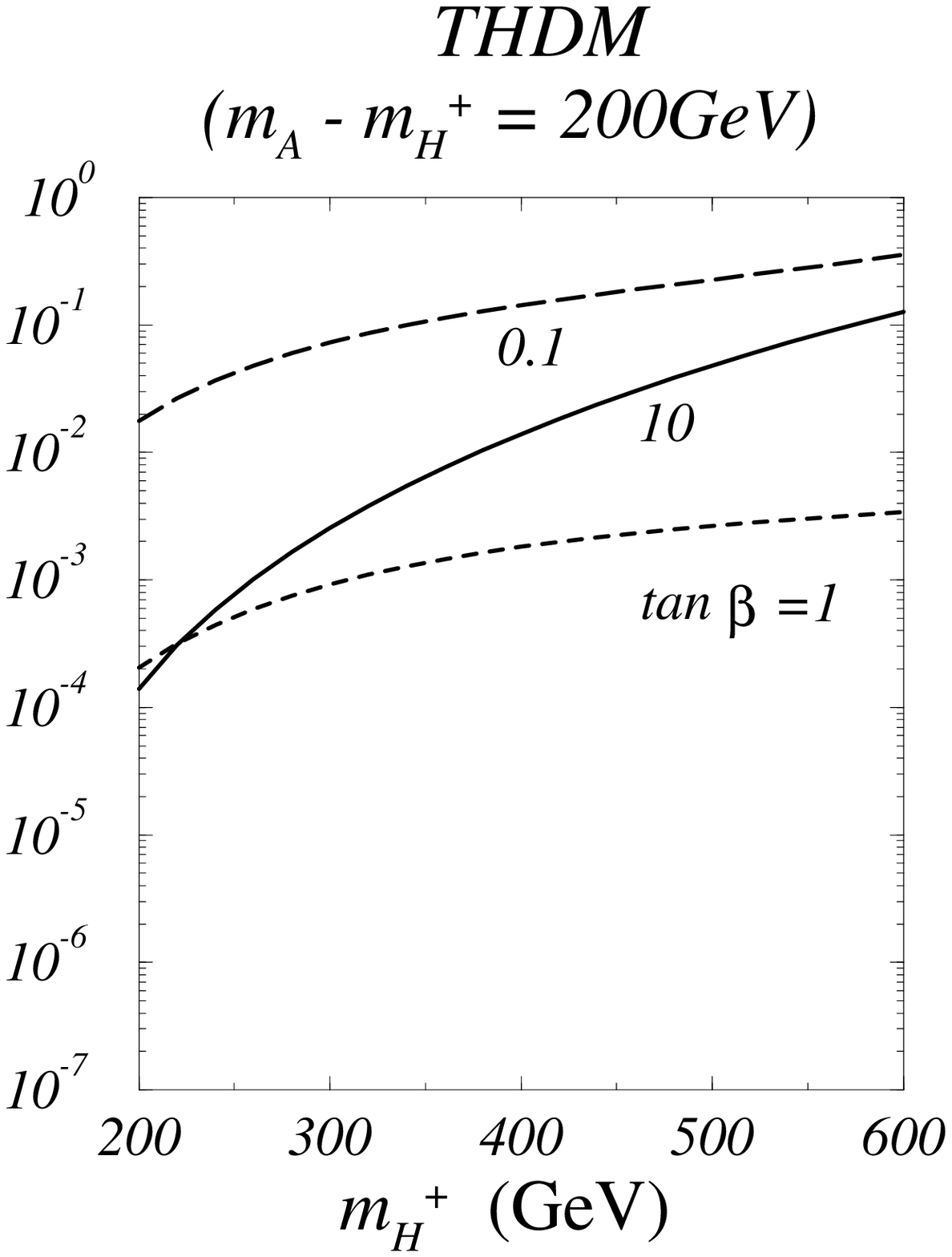}
  {\bf Fig 4(b):} $\Gamma(H^+\rightarrow W^+Z^0)$
                  for $\tan\be =$ 0.1, 1 and 10 in THDM with 
                  $m_{A^0}- m_{H^\pm} = 200$ GeV. 
                  The other parameters are taken as $\al = \be - \pi/2$, 
                  $m_{H^0} = m_{H^\pm} + 10$ GeV, $m_{h^0} = 140$ GeV 
                  and $\mu_3 = 0$.
\end{minipage}
\begin{minipage}[t]{7cm}
\epsfxsize=8cm
\epsfbox{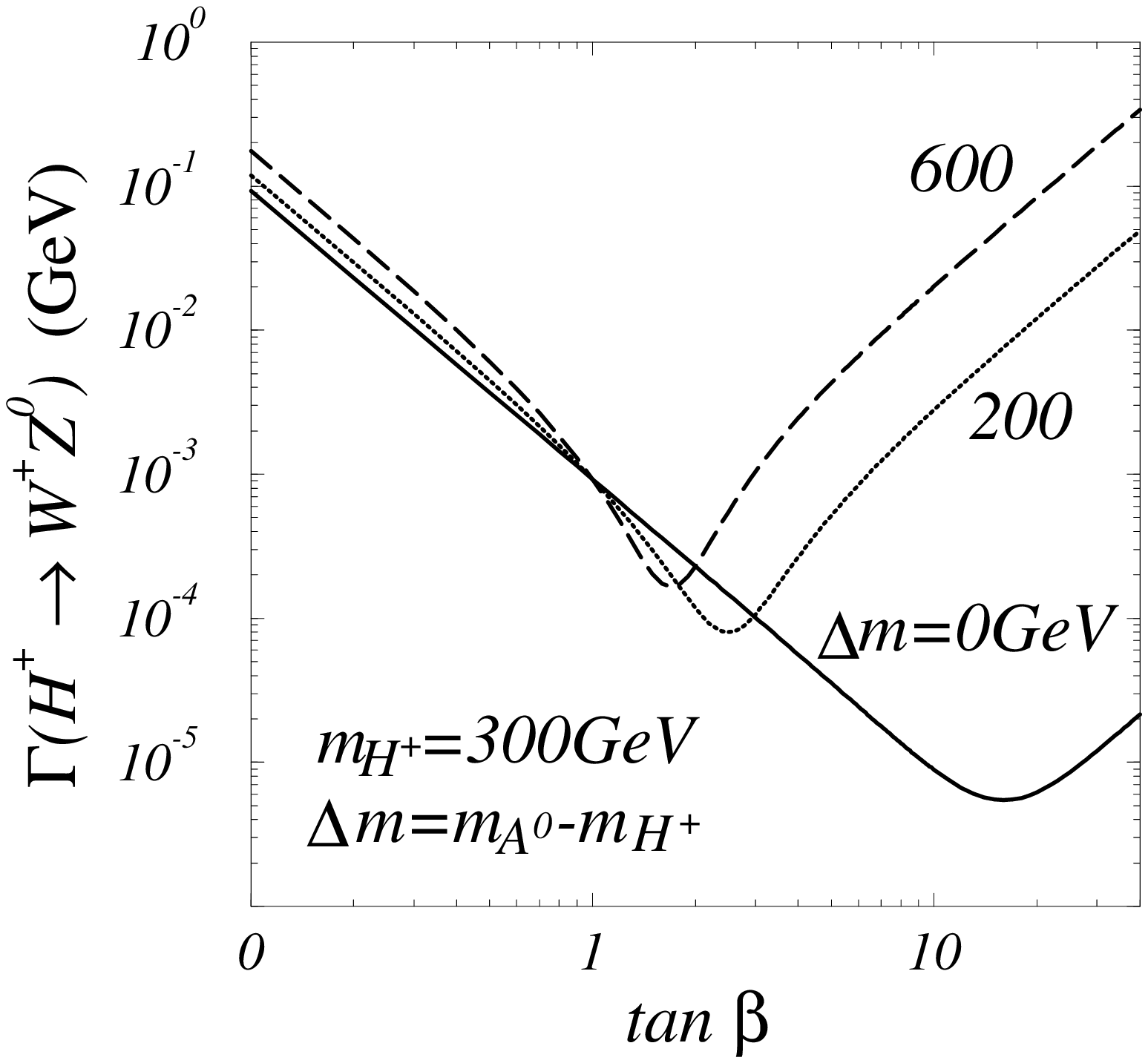}
  {\bf Fig 5:} $\Gamma(H^+\rightarrow W^+Z^0)$ at $m_{H^\pm} = 300$ GeV 
                  as a function of  $\tan\be$. 
                  $\Delta m = m_{A^0}- m_{H^\pm}$ is set into 
                  $0$, $200$ or $600$ GeV. 
                  The other parameters are taken as $\al = \be - \pi/2$, 
                  $m_{H^0} = 310$ GeV, $m_{h^0} = 140$ GeV and $\mu_3 = 0$.
\end{minipage}

\begin{minipage}[t]{7cm}  
\epsfxsize=8.5cm
\epsfbox{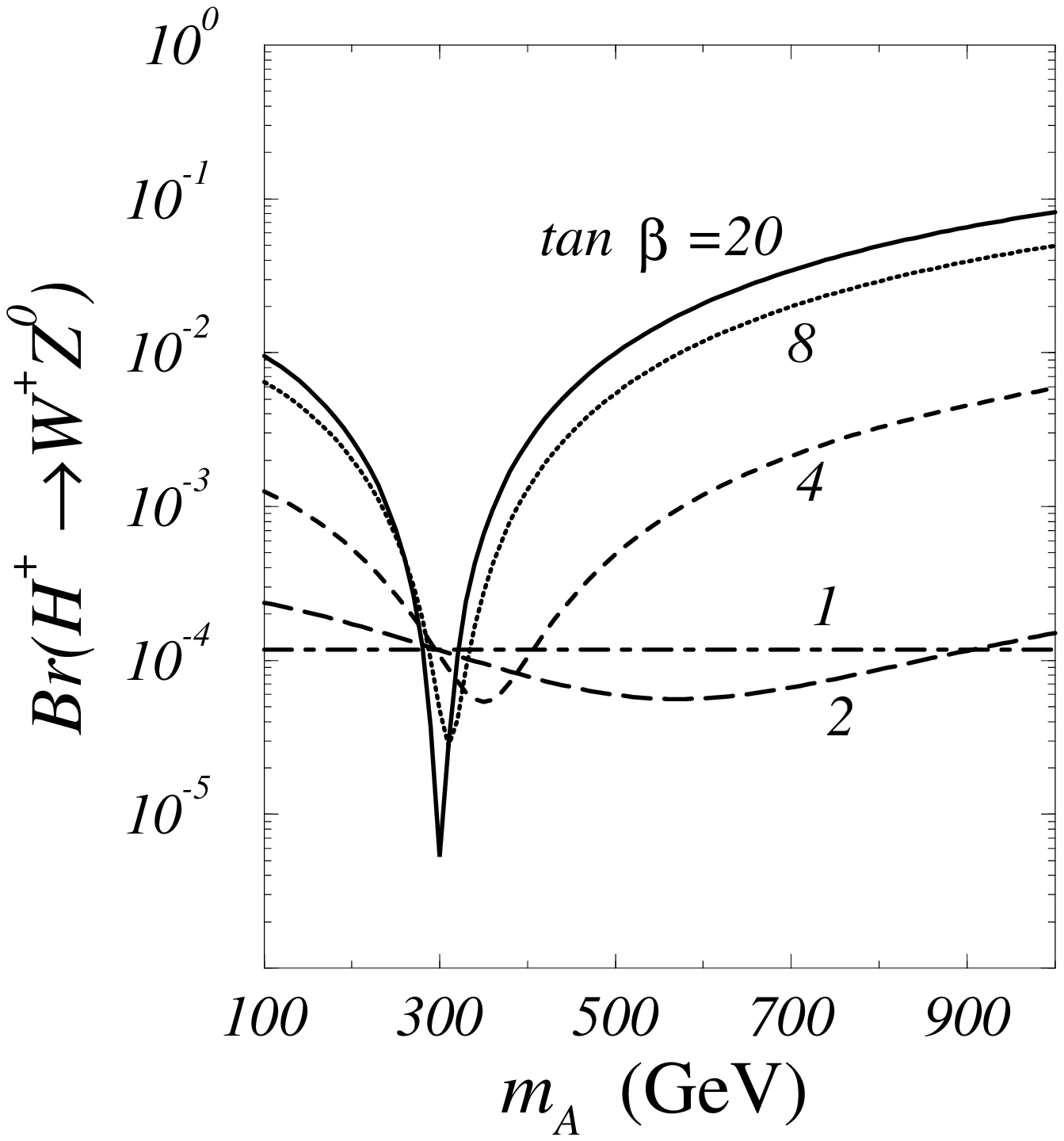}
  {\bf Fig 6(a):} $Br(H^+ \rightarrow W^+ Z^0)$ in THDM 
                  for $m_{H^\pm} = 300$ GeV 
                  as a function of $m_{A^0}$.   
                  Other parameters are chosen as $\al = \be - \pi/2$, 
                  $m_{H^\pm} = 310$ GeV, $m_{h^0} = 140$ GeV and 
                  $\mu_3 = 0$.
\end{minipage}
\begin{minipage}[t]{7cm}
\epsfxsize=8.5cm
\epsfbox{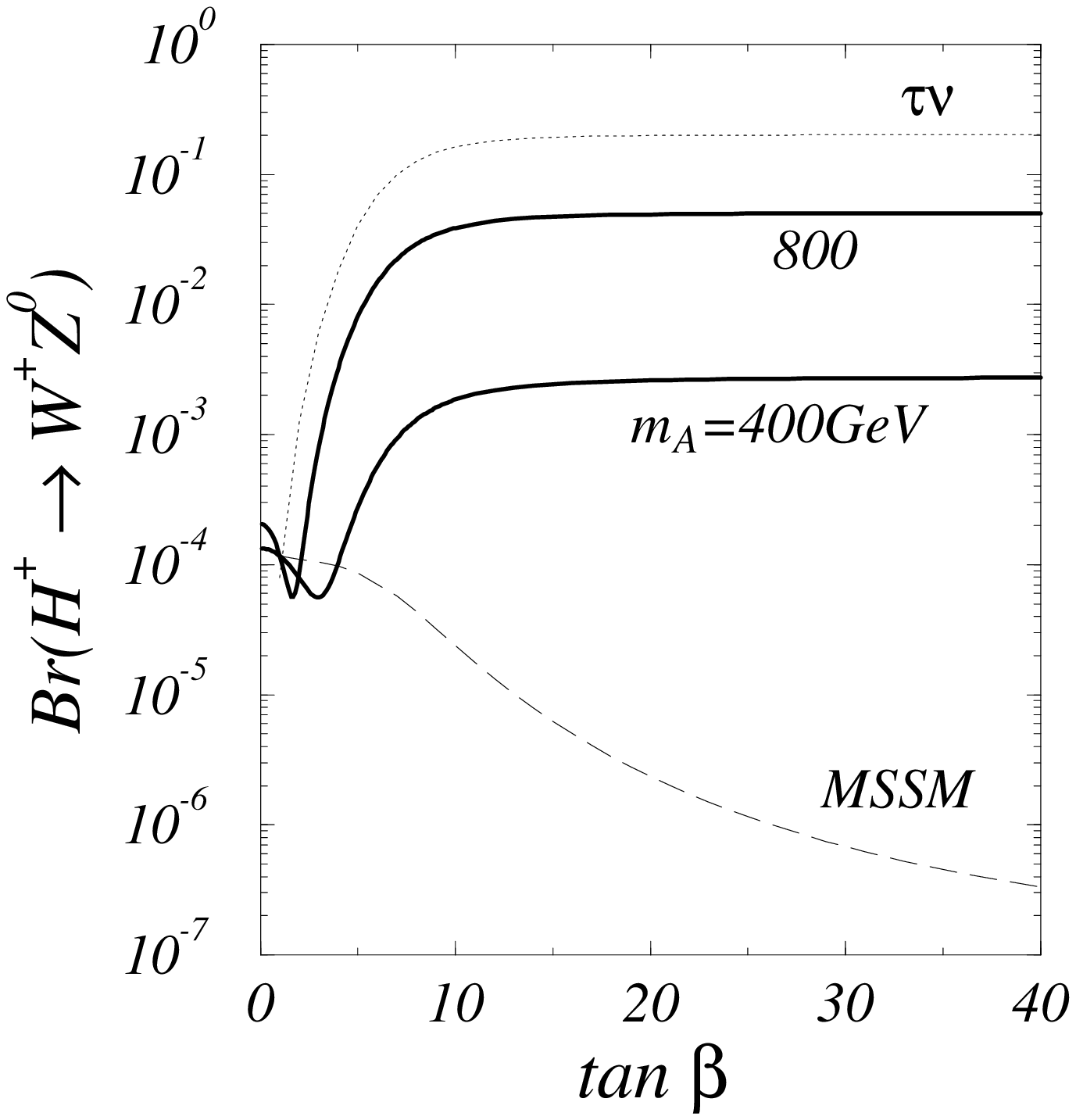}
  {\bf Fig 6(b):}  $Br(H^+ \rightarrow W^+ Z^0)$ in THDM 
                   for $m_{H^\pm} = 300$ GeV
                   as a function of $\tan \be$ (Solid lines).  
                   Other parameters are chosen as the same  
                   as Fig 6(a). 
\end{minipage}

\vspace*{1cm}
\noindent
 $\sim 175$ GeV  \cite{hmh,top}. 
In addition, there is a lower bound for the charged Higgs boson 
in Type II THDM
from the $b \rightarrow s \gamma$ measurement as
$m_{H^\pm} > 244 + 63/(\tan \be)^{1.3}$ \cite{cleo}.  
We assume here that the charged Higgs boson is heavy enough to 
open the $tb$ mode, which becomes the most dominant mode then.   
The other modes to be considered are 
$H^\pm \rightarrow \tau \nu, \; cs$ and $h^0W^\pm$ (if it is allowed).
Their decay widths at tree level behave like 
\begin{eqnarray}
  \Gamma (H^\pm \rightarrow tb) &\sim& 
       N_C (m_t^2 \cot^2 \beta + m_b^2 \tan^2 \beta), \\
  \Gamma (H^\pm \rightarrow cs) &\sim& 
       N_C (m_c^2 \cot^2 \beta + m_s^2 \tan^2 \beta), \\
  \Gamma (H^\pm \rightarrow \tau \nu) 
       &\sim& m_\tau^2 \tan^2 \beta,   \\
  \Gamma (H^\pm \rightarrow h^0 W^\pm) 
       &\sim& \mw  \cos^2(\al - \be), 
\end{eqnarray}
The region of variables are taken as 
$200 < m_{H^\pm} < 800$ GeV, $100 < m_{A^0} < 900$ GeV and 
$1 < \tan \be < 50$.  
The other parameters are fixed as $m_{h^0} = 140$ GeV, 
$m_{H^0} = 310$ GeV and $\al = \be - \pi/2$. 
As to the fermion masses, 
we assume that $m_t = 175$ GeV and 
$m_b (m_{H^\pm}) = 3$ GeV \cite{kiyo}.  
The reason for this parameter choice is the $\rho$-parameter constraint. 
From the data from LEP experiments, we can evaluate the contribution of 
the Higgs sector; $\Delta \rho_{THDM}$ \\
\vspace{1cm}
\begin{minipage}[t]{7.5cm}  
\epsfxsize=8.5cm
\epsfbox{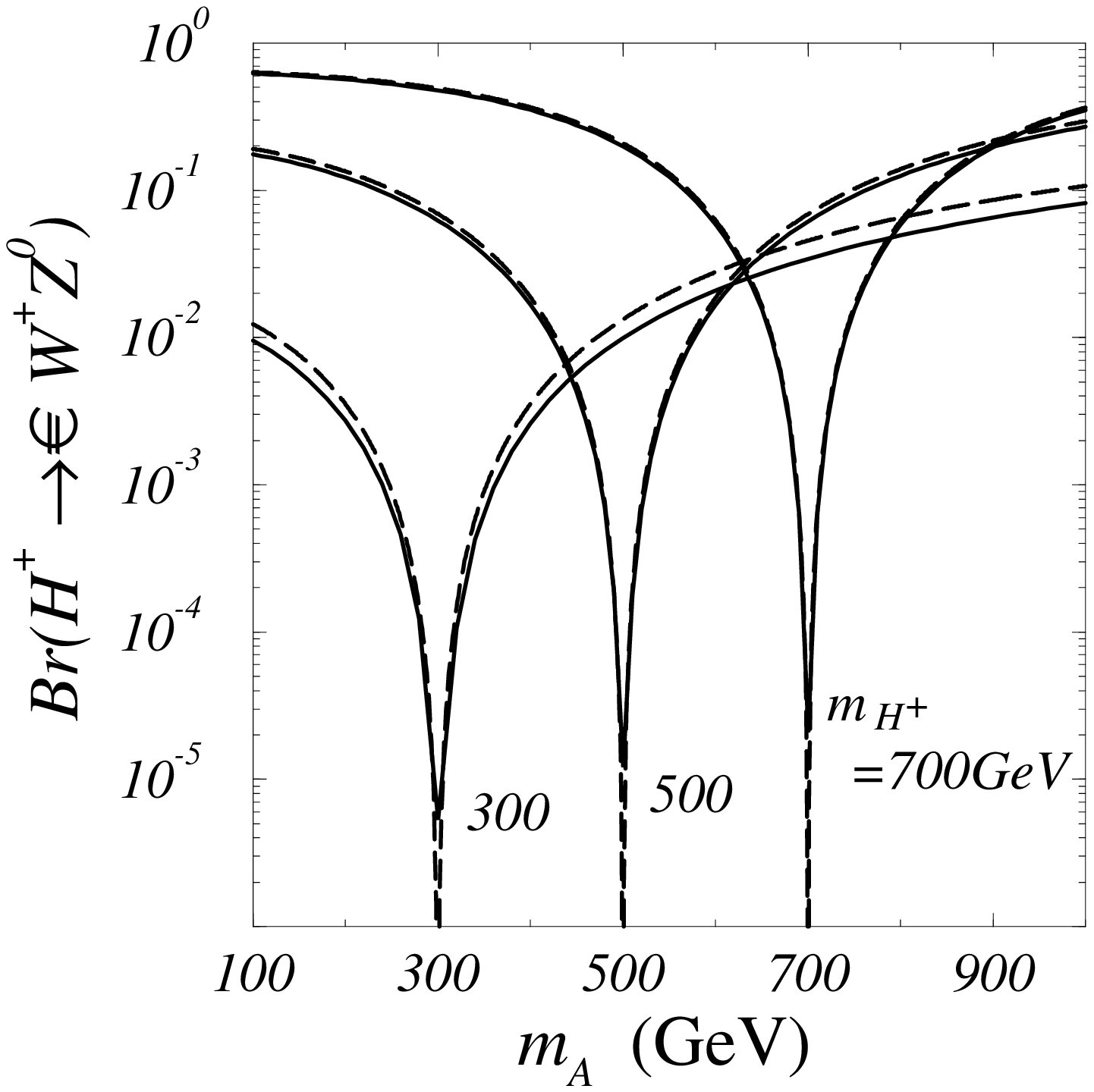}
  {\bf Fig 7(a):}  $Br(H^+ \rightarrow W^+ Z^0)$ for various values of 
                   $m_{H^\pm}$ as a function of $m_{A^0}$.  
                   The other parameters are chosen as $\tan \be = 20$, 
                   $\al = \be - \pi/2$, $m_{H^0} = m_{H^\pm} +10$ GeV, 
                   $m_{h^0} = 140$ GeV and $\mu_3 = 0$. 
                   Solid lines are the results of the full calculation. 
                   Dashed lines are those of the calculation 
                   by using the equivalence theorem.      
\end{minipage}
\begin{minipage}[t]{7.5cm}
\epsfxsize=8.5cm
\epsfbox{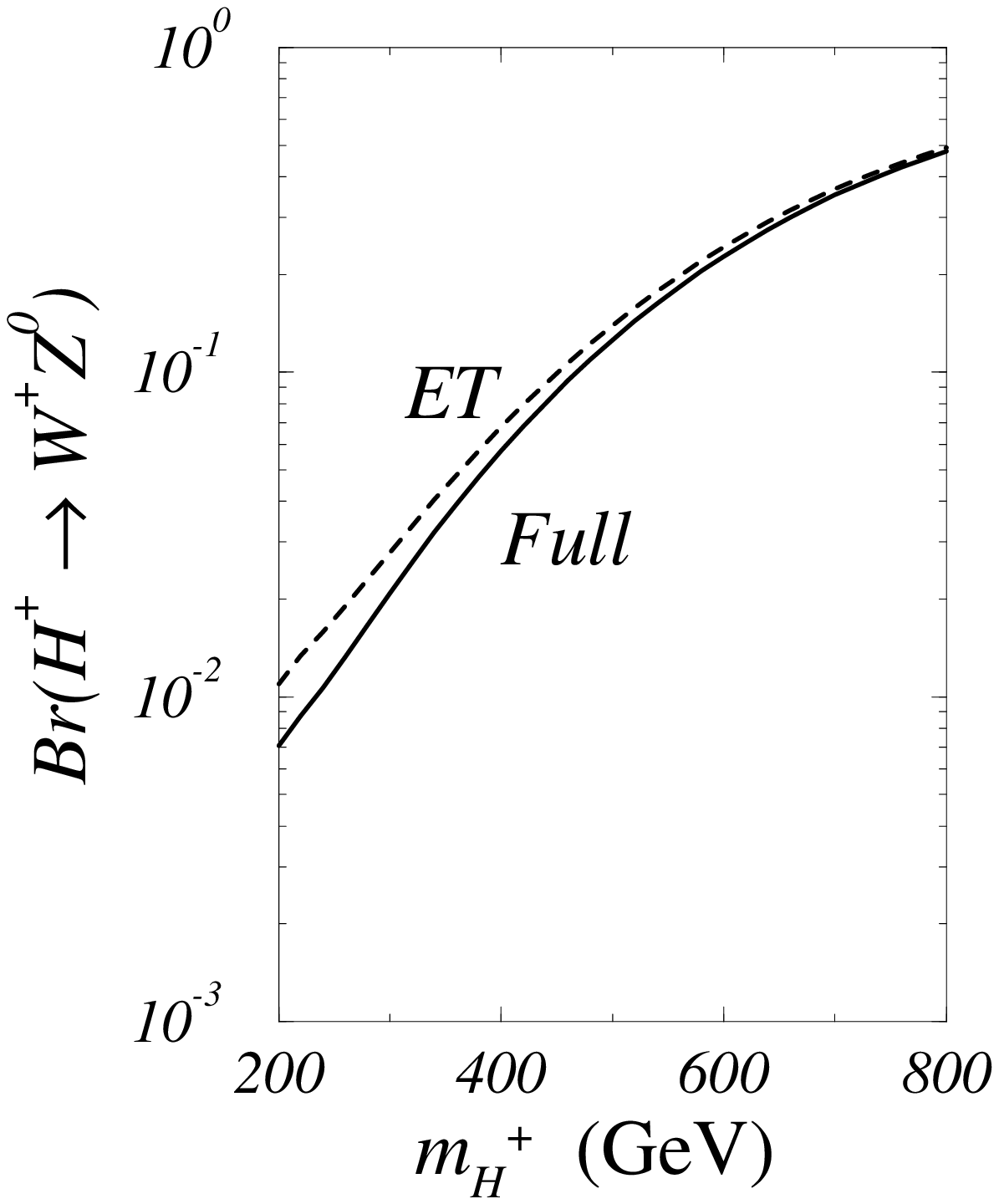}
  {\bf Fig 7(b):}  $Br(H^+ \rightarrow W^+ Z^0)$ for $\tan \be = 20$.  
                   Other parameters are chosen as $\tan \be = 20$, 
                   $\al = \be - \pi/2$, 
                   $m_{A^0} = m_{H^\pm} +300$ GeV, 
                   $m_{H^0} = m_{H^\pm} +10$ GeV,                   
                   $m_{h^0} = 140$ GeV and $\mu_3 = 0$. 
                   The results from the full calculation (Full) and 
                   the results by using the equivalence theorem (ET)
                   are shown.
\end{minipage}\\

\noindent
$\sim - 0.00180 \pm 0.00204$ ($2 \sigma$) \cite{hmh}.  
The other constraints for $\tan \be$ from $B^0-{\bar B^0}$ mixing \cite{bb} 
and for $m_{H^\pm}$ from $b \rightarrow s \gamma$ \cite{cleo} are also in 
our consideration here. Moreover, we take account of the constraint 
from the perturbative unitarity for the Higgs boson masses \cite{uni}.

In Fig 6, we show the branching ratio $Br(H^\pm \rightarrow W^\pm Z^0)$ 
at $m_{H^\pm}=300$ GeV. 
We can see in Fig 6(a) 
that the branching ratio become larger than $10^{-2}$ 
if the mass splitting between $H^\pm$ and $A^0$ is greater than 200 GeV 
for $\tan \be > 5 \sim 8$.  
The maximal value of $Br(H^\pm \rightarrow W^\pm Z^0)$ 
can amount to near $10^{-1}$ for very large $m_{A^0}$ and 
$\tan \beta > 20$. 
In the nearly $SU(2)_V$ symmetric cases in the Higgs sector 
($m_{A^0} \sim m_{H^\pm}$), the Higgs non-decoupling effects 
are canceled out and only the fermion and gauge boson contributions remain, 
so that the branching ratio becomes smaller than $10^{-4}$. 
Since the top-quark mass contributions are decreased and the Higgs mass 
contributions are increased as $\tan \be$ grows, the cancellation 
of the Higgs contributions become more clear for large $\tan \be$. 
On the other hand, at near $\tan \be \sim 1$, the top-quark mass 
contribution becomes much dominant and thus we can see nothing happens 
at $m_{H^\pm} = m_{A^0}$. 
The $\tan \be$ dependence is shown in Fig 6(b) for $m_{H^\pm} = 300$ GeV.  
The other parameters are taken as same as Fig 6(a).    
We also show there $Br(H^+ \rightarrow W^+Z^0)$ in the MSSM case   
for a comparison, 
in which the Higgs mass effects is almost suppressed by the 
approximate $SU(2)_V$ in the Higgs sector. 
For further  comparison, the results of 
$Br(H^+ \rightarrow \tau \nu)$ are also attached in Fig 6(b). 
The branching ratios for various values of $m_{H^\pm}$ 
are shown in Fig 7(a) and (b).
The similar properties to $m_{H^\pm} \sim 300$ GeV 
are seen for each value of $m_{H^\pm}$ in Fig 7(a). 
Both the results by the full calculation and the calculation 
simplified by virtue of the equivalence theorem are presented there.
We can see that the latter results become an excellent approximation 
to the full calculation for larger values of $m_{H^\pm}$ especially 
in Fig 7(b).

We have shown that, in general THDM, 
the loop induced $H^\pm W^\mp Z^0$ vertex 
can be considerably enhanced due to the non-decoupling effects 
of the Higgs bosons.  
All the analyses above have been considered by making 
the soft-breaking parameter $\mu_3^2$ to be zero 
because we have tried to extract the Higgs non-decoupling effects 
as large as possible.
However, the soft-breaking parameter $\mu_3^2$ often become very 
important in various aspects of physics.  
First of all, it cannot be neglected in MSSM case.   
Second, if we are interested  
in the Higgs sector as 
an additional CP violating source, $\mu_3^2$ is the parameter 
which can be considered to have a phase \cite{cp}.  
Finally, since $\mu_3^2 = 0$ implies that 
there is the exact 
discrete symmetry in the Higgs sector 
which is spontaneously broken according to the gauge symmetry breaking.
In that case, the problem of the domain wall takes place \cite{dom}.
To avoid this, the discrete symmetry may have to be explicitly broken
(but only softly for FCNC suppression) by the non-zero $\mu_3^2$ parameter. 
As we mentioned in Sec 3, the heavier Higgs boson masses 
have two kinds of origin, namely, the quartic coupling 
constant times the vacuum expectation value and the soft-breaking term. 
The non-zero soft-breaking term reduces the contribution of 
the quartic couplings for a fixed masses, 
so that the non-decoupling effects are suppressed to some extent.
We show the relation between the non-decoupling effects and 
the non-zero soft-breaking parameter on the branching ratio in Fig 8. 
The  Higgs\\
\begin{center}
\begin{minipage}[t]{10cm}  
\epsfxsize=8.5cm
\epsfbox{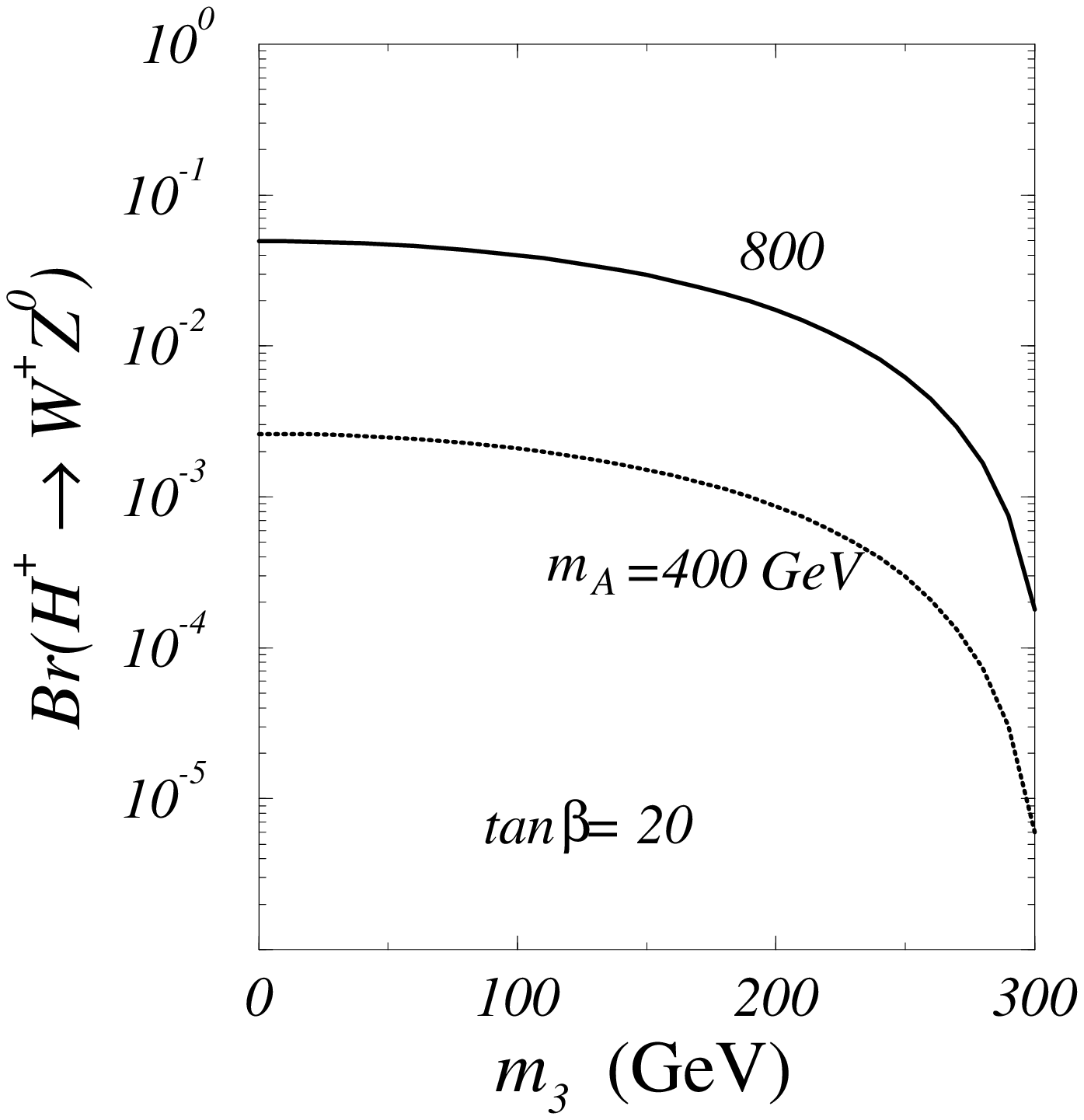}
  {\bf Fig 8:}  The soft-breaking parameter dependence of  
                $Br(H^+ \rightarrow W^+ Z^0)$ 
                at $m_{H^\pm} = 300$ GeV.  
                The parameter $m_3$ is defined by 
                $m_3^2 \equiv \mu_3^2/\cos \be \sin \be$.  
                Other parameters are chosen as $\al = \be - \pi/2$, 
                $m_{H^0} = 310$ GeV and $m_{h^0} = 140$ GeV. 
\end{minipage}
\end{center}
\vspace{5mm}

\noindent
non-decoupling limit 
($m_3^2 \equiv \mu_3^2/\sin \be \cos \be \sim 0$) 
is what we have seen in the previous figures. 
In the case of $m_3 \sim 300$ GeV ($= m_{H^\pm}$), in which 
the heavier Higgs bosons receive 
their masses only from $\mu_3^2$, 
the complete cancellation of the non-decoupling effects 
of Higgs boson masses takes place and 
the model becomes the decoupling theory 
for the heavy Higgs bosons just like in MSSM.


\section{Discussion and Conclusion}

\hspace*{18pt}
We have discussed the loop-induced $H^\pm W^\mp Z^0$ vertex in Type II 
THDM and its enhancement due to the non-decoupling effects of the heavy 
Higgs as well as the heavy quarks. 
The conditions for the large enhancement due to the Higgs mass effects  
have been summarized as 
1) the large Higgs masses coming from the larger 
   contributions of the quartic coupling constants 
   with keeping the soft-breaking parameter to be smaller, 
2) the large explicit breaking of $SU(2)_V$ in the Higgs sector 
   by the large mass splitting between $H^\pm$ and $A^0$.    
In the case of MSSM, since both the conditions above are impossible, 
such the Higgs mass effects do not take place and
any substantial enhancement cannot be realized. 
The decay width $\Gamma(H^+ \rightarrow W^+ Z^0)$ remains $< 10^{-3}$ GeV 
at $m_{m^\pm} = 300$ GeV for $\tan \be >1$. 
On the other hand, in the case of THDM, the conditions can be satisfied 
within the constraint from the available experimental data.      
We have found that the decay width $\Gamma (H^\pm \rightarrow W^+ Z^0)$ 
can amount to $> 10^{-1}$ GeV at $m_{H^\pm} \sim 300$ GeV 
for large $\tan \be$.
These values are considered to be much larger ($\sim 10^1 - 10^2$) than the 
typical values in the $E_6$ model,  
in which a tree $H^\pm W^\mp Z^0$ coupling can 
be induced through $Z-Z'$ mixing \cite{e6} 
and also smaller ($\sim 10^{-2} - 10^{-1}$) 
than those in the model with a doublet and two (a real and a complex) 
triplets \cite{exo}. 
In Type II THDM, since $m_{H^\pm}$ is constrained to be larger than 
$\sim 250$ GeV for large $\tan \be$ by $b \rightarrow s \gamma$ data, 
$H^\pm \rightarrow tb$ opens and becomes the most dominant mode. 
We have evaluated the branching ratio and found that,  
even in such the situation, it can amount to 
$10^{-2} \sim 10^{-1}$ at $m_{H^\pm} = 300$ GeV  
within the constraints from the present experimental data and also  
the perturbative unitarity. 

Such the enhancement may make it possible to detect the decay mode 
at LHC or (if fortune) LC's.
The charged Higgs boson is mainly produced through the subprocess 
$gb \rightarrow t H^\pm$ \cite{bhs} at LHC.   
We expect that, if $m_{H^\pm}$ is $300$ GeV, about 200 (50) events of 
$H^\pm \rightarrow W^\pm Z^0 \rightarrow lll\nu$ are produced 
for $\tan \be = 1$ (20) per a year 
at LHC with the integrated luminosity $\sim 2 \times 10^2$ fb$^{-1}$/year.  
Since the background (mainly $ud \rightarrow W^\pm Z^0$) has been naively 
estimated to be such that a few \% of the branching ratios are required 
to see a signal, we can expect to detect the decay mode if such 
the large enhancement occurs.      
At the future $e^+e^-$ linear collider with $\sqrt{s} = 1$ TeV and 
the integrated luminosity $\sim$ 50 fb$^{-1}$/year, 
a few thousands of $H^\pm$ are expected to be produced through 
$e^+e^- \rightarrow H^+H^-$.         
The decay $H^\pm \rightarrow W^\pm Z^0$ with the maximally 
enhanced branching ratio (near 10 \%) 
might be also detectable there because less background would be expected.

\vspace{6mm}
\noindent
{\large \em Acknowledgments}

The author would like to thank Yasuhiro Okada and Kaoru Hagiwara for valuable 
discussions and useful comments, Chung Kao, Hide-Aki Tohyama, Minoru Tanaka  
and Seiji Matsumoto for helpful discussions. 
This work was supported, in part, by Grant-in-Aid for Scientific Research 
from the Ministry of Education, Science and Culture of Japan.

\newpage

\appendix
\renewcommand{\theequation}{\thesection.\arabic{equation}}
\setcounter{section}{1}
\addcontentsline{toc}{section}{APPENDICES}
\begin{center}
{\large \bf APPENDICES}
\end{center}
\setcounter{equation}{0}
\subsection*{APPENDIX 1: Calculation of Boson-Loop Contribution}

\hspace*{18pt}
Here we show the explicit results of the calculation of the loop induced 
$H^\pm W^\mp Z^0$ vertex.
As shown in Eq \eq{ff}, the contributions
 are expressed 
in terms of the quantities $F$, $G$ and $H$. 
They can be divided as 
\begin{eqnarray}
  X = X^{({\rm a})} + X^{({\rm b})} + X^{({\rm c})} + X^{({\rm d})},
\end{eqnarray}
where $X$ represents $F$, $G$ or $H$. 
$X^{({\rm a} \sim {\rm d}) }$  correspond to  
the diagrams in Fig 2(a) $\sim$ (d), respectively.   
We employ the 't Hooft-Feynman gauge for calculation of the boson-loop
diagrams. Explicit forms for $X^{({\rm a}) \sim ({\rm c})}$ are listed in the 
following. 
The contribution of fermion-loop diagrams $X^{\rm (d)}$ have been 
calculated in the unitary gauge 
and their explicit expressions are given in Ref \cite{capd}.
As mentioned before, the diagrams with a fermion loop, 
by themselves, form the gauge invariant sub-set, so that 
we can use these results consistently.     
The contributions of the boson-loop diagrams 
are expressed in terms of the integral functions \cite{pave}, whose  
definition here is beased on the second paper of Ref \cite{hmh}.    
\begin{eqnarray}
&& \!\!\!\!\!\!\!\!\!\!\!\! 
  F^{\rm (a)} = \fr{2}{16 \pi^2 v^2 \cos \theta_W} \nn \\ 
&& \!\!\!\!\!\!\!\!\!\!\!\! 
   \times \lt[  - \lt\{ K(\al,\be) \mH  + 
             J(\al,\be) (- \mH + 2 \mg) 
             - \fr{\sin (\al + \be)}
                  {\sin \be \cos \be} m_3^2 \rt\}  C_{24}[H^\pm A^0H^0] 
\rt.  \nn \\
&& \!\!\!\!\!\!\!\!\!\!\!\! 
     - \lt\{ L(\al,\be) \mh  - 
             J(\al,\be) (- \mh + 2 \mg) 
             - \fr{\cos (\al + \be)}
                  {\sin \be \cos \be} m_3^2 \rt\}  C_{24}[H^\pm A^0h^0] \nn \\
&& \!\!\!\!\!\!\!\!\!\!\!\! 
     + \lt\{ K(\al,\be) \mH  + 
             J(\al,\be) (- \mH + 2 \mg) 
             - \fr{\sin (\al + \be)}
                  {\sin \be \cos \be} m_3^2 \rt\}  
           \cs 2\theta_W C_{24}[H^0H^\pm H^\pm] \nn \\
&& \!\!\!\!\!\!\!\!\!\!\!\! 
     + \lt\{ L(\al,\be) \mh  - 
             J(\al,\be) (- \mh + 2 \mg) 
             - \fr{\cos (\al + \be)}
                  {\sin \be \cos \be} m_3^2\rt\}  
           \cs 2\theta_W C_{24}[h^0H^\pm H^\pm] \nn \\
&& \!\!\!\!\!\!\!\!\!\!\!\! 
     + J(\al,\be) (\mg - \mH) C_{24}[w^\pm z^0 H^0]
     - J(\al,\be) (\mg - \mh) C_{24}[w^\pm z^0 h^0] \nn \\
&& \!\!\!\!\!\!\!\!\!\!\!\! 
     - J(\al,\be) (\mg - \mH) \cs 2\theta_W C_{24}[H^0 w^\pm w^\pm] \nn \\
&& \!\!\!\!\!\!\!\!\!\!\!\! 
     + J(\al,\be) (\mg - \mh) \cs 2\theta_W C_{24}[h^0 w^\pm w^\pm] \nn \\
&& \!\!\!\!\!\!\!\!\!\!\!\! 
     - J(\al,\be) (\mg - \ma) 
     \lt( C_{24}[w^\pm H^0 A^0] -
      C_{24}[w^\pm h^0 A^0] \rt) \nn \\
&& \!\!\!\!\!\!\!\!\!\!\!\!  
     - J(\al,\be) \mw 
      \lt( C_{24} [W^\pm H^0 A^0] - 
           C_{24} [W^\pm h^0 A^0] \rt) \nn \\
&& \!\!\!\!\!\!\!\!\!\!\!\!  
     + J(\al,\be) \fr{\cs 2\theta_W}{\cs \theta_W} 
       m_W^2 C_{24} [H^\pm H^0 Z^0] 
     - J(\al,\be) \fr{\cs 2\theta_W}{\cs \theta_W} 
       m_W^2 C_{24} [H^\pm h^0 Z^0] \nn \\
&& \!\!\!\!\!\!\!\!\!\!\!\!  
     - J(\al,\be) \mw
     \lt\{ 4 (\mw + p_W \cdot p_Z ) C_0 
     + 2 (2 p_W + p_Z) \cdot (p_W C_{11} + p_Z C_{12}) \rt.\nn\\
      && \;\;\;\;\;\;\;\;\;\;\;\;\;\;\;\;\;\;  
      \lt. + p_W \cdot p_Z C_{23} + 4 C_{24} \rt\}
      [W^\pm Z^0 H^0] \nn\\
&& \!\!\!\!\!\!\!\!\!\!\!\!  
     + J(\al,\be) \mw
     \lt\{ 4 (\mw + p_W \cdot p_Z ) C_0 
     + 2 (2 p_W + p_Z) \cdot (p_W C_{11} + p_Z C_{12}) \rt.\nn\\
      && \;\;\;\;\;\;\;\;\;\;\;\;\;\;\;\;\;\;  
      \lt. + p_W \cdot p_Z C_{23} + 4 C_{24} \rt\}
      [W^\pm Z^0 h^0] \nn\\
&& \!\!\!\!\!\!\!\!\!\!\!\!  
     + J(\al,\be) \cs^2 \theta_W \mw 
     \lt\{ (\mz - \mw ) C_0 
     - 2 p_Z \cdot (p_W C_{11} + p_Z C_{12}) \rt.\nn\\
      && \;\;\;\;\;\;\;\;\;\;\;\;\;\;\;\;\;\;  
      \lt. + p_W \cdot p_Z C_{23} + 4 C_{24} \rt\}
      [H^0 W^\pm W^\pm] \nn\\
&& \!\!\!\!\!\!\!\!\!\!\!\!  
     - J(\al,\be) \cs^2 \theta_W \mw 
     \lt\{ (\mz - \mw ) C_0 
     - 2 p_Z \cdot (p_W C_{11} + p_Z C_{12}) \rt.\nn\\
      && \;\;\;\;\;\;\;\;\;\;\;\;\;\;\;\;\;\;  
      \lt. + p_W \cdot p_Z C_{23} + 4 C_{24} \rt\}
      [h^0 W^\pm W^\pm] \nn\\
&& \!\!\!\!\!\!\!\!\!\!\!\! 
     - J(\al,\be) \mz (\mg - \mH) \sn^2 \theta_W C_0[w^\pm Z^0 H^0] \nn \\
&& \!\!\!\!\!\!\!\!\!\!\!\! 
     + J(\al,\be) \mz (\mg - \mh) \sn^2 \theta_W C_0[w^\pm Z^0 h^0] \nn \\
&& \!\!\!\!\!\!\!\!\!\!\!\! 
     - J(\al,\be) \mw (\mg - \mH) \sn^2 \theta_W C_0[H^0 W^\pm w^\pm] \nn \\
&& \!\!\!\!\!\!\!\!\!\!\!\! 
     + J(\al,\be) \mw (\mg - \mh) \sn^2 \theta_W C_0[h^0 W^\pm w^\pm] \nn \\
&& \!\!\!\!\!\!\!\!\!\!\!\! 
   \lt.  + J(\al,\be) \mw \sn^2 \theta_W 
       \lt( C_{24}[H^0 w^\pm W^\pm] 
     -  C_{24}[h^0 w^\pm W^\pm]\rt) \rt],
\end{eqnarray}
where $m_3$ is defined as $m_3^2 = \mu_3^2/\sin \be \cos \be$. 
\begin{eqnarray}
&& \!\!\!\!\!\!\!\!\!\!\!\! 
      F^{\rm (b)} = \fr{2}{16 \pi^2 v^2 \cos \theta_W} \nn \\ 
&& \!\!\!\!\!\!\!\!\!\!\!\! 
    \times \lt[ - \fr{1}{2} \lt\{ K(\al,\be) \mH  + 
          J(\al,\be) (- \mH + 2 \mg) 
             - \fr{\sin (\al + \be)}
                  {\sin \be \cos \be} m_3^2\rt\} \sn^2 \theta_W 
          B_0 [H^0 H^\pm] \rt.\nn \\
&& \!\!\!\!\!\!\!\!\!\!\!\! 
     - \fr{1}{2} \lt\{ L(\al,\be) \mh  - 
             J(\al,\be) (- \mh + 2 \mg) 
             - \fr{\cos (\al + \be)}
                  {\sin \be \cos \be} m_3^2\rt\} \sn^2 \theta_W
          B_0 [h^0 H^\pm] \nn \\
&& \!\!\!\!\!\!\!\!\!\!\!\! 
     - \fr{1}{2} J(\al,\be) (\mg - \mH) \sn^2 \theta_W 
          B_0 [H^0 w^\pm]  
     + \fr{1}{2} J(\al,\be) (\mg - \mh) \sn^2 \theta_W
          B_0 [h^0 w^\pm] \nn \\
&& \!\!\!\!\!\!\!\!\!\!\!\! 
     +  J(\al,\be) \mw \sn^2 \theta_W
        \lt(  B_0 [p_W; W^\pm H^0] 
     -    B_0 [p_W; W^\pm h^0] \rt) \nn \\
&& \!\!\!\!\!\!\!\!\!\!\!\!
     +  J(\al,\be) \mz \sn^2 \theta_W
        \lt(  B_0 [p_Z; Z^0 H^0] 
     -    B_0 [p_Z; Z^0 h^0] \rt) \nn \\
&& \!\!\!\!\!\!\!\!\!\!\!\! 
     + \fr{1}{2}\lt\{ K(\al,\be) \mH + 
                      J(\al,\be) ( - \mH + 2 \mg) 
             - \fr{\sin (\al + \be)}
                  {\sin \be \cos \be} m_3^2\rt\} 
      \fr{\mH - \mg}{p^2 - \mw} \sn^2 \theta_W
        B_0 [H^0 H^\pm]  \nn \\
&& \!\!\!\!\!\!\!\!\!\!\!\! 
     + \fr{1}{2}\lt\{ L(\al,\be) \mh - 
                      J(\al,\be) ( - \mh + 2 \mg) 
             - \fr{\cos (\al + \be)}
                  {\sin \be \cos \be} m_3^2\rt\} 
       \fr{\mh - \mg}{p^2 - \mw} \sn^2 \theta_W
        B_0 [h^0 H^\pm]  \nn \\
&& \!\!\!\!\!\!\!\!\!\!\!\! 
     + \fr{1}{2} J(\al,\be) \mH  
       \fr{\mH - \mg}{p^2 - \mw} \sn^2 \theta_W
        B_0 [H^0 w^\pm]  \nn \\
&& \!\!\!\!\!\!\!\!\!\!\!\! 
     - \fr{1}{2} J(\al,\be) \mh  
       \fr{\mh - \mg}{p^2 - \mw} \sn^2 \theta_W
        B_0 [h^0 w^\pm]  \nn \\
&& \!\!\!\!\!\!\!\!\!\!\!\! 
     + \fr{1}{2} \lt\{ K(\al,\be) \mH +  
                       J(\al,\be) ( - \mH + 2 \mg) 
             - \fr{\sin (\al + \be)}
                  {\sin \be \cos \be} m_3^2\rt\} 
       \nn\\
&&  \times \fr{\mw}{p^2 - \mw} \sn^2 \theta_W
        (B_0 + 2 B_1) [H^0 H^\pm]  \nn \\
&& \!\!\!\!\!\!\!\!\!\!\!\! 
     + \fr{1}{2} \lt\{ L(\al,\be) \mh -  
                       J(\al,\be) ( - \mh + 2 \mg) 
             - \fr{\cos (\al + \be)}
                  {\sin \be \cos \be} m_3^2\rt\} 
       \nn\\
&& \times \fr{\mw}{p^2 - \mw} \sn^2 \theta_W
        (B_0 + 2 B_1) [h^0 H^\pm]  \nn \\
&& \!\!\!\!\!\!\!\!\!\!\!\! 
     + J(\al,\be) m_W^4  
       \fr{1}{p^2 - \mw} \sn^2 \theta_W
        (B_0 - B_1) [H^0 W^\pm]  \nn \\
&& \!\!\!\!\!\!\!\!\!\!\!\! 
     - J(\al,\be) m_W^4  
       \fr{1}{p^2 - \mw} \sn^2 \theta_W
        (B_0 - B_1) [h^0 W^\pm]  \nn \\
&& \!\!\!\!\!\!\!\!\!\!\!\! 
     + J(\al,\be) \mw  
       \fr{\mH - \mg}{p^2 - \mw} \sn \theta_W
        (B_0 + 2 B_1) [H^0 w^\pm]  \nn \\
&& \!\!\!\!\!\!\!\!\!\!\!\! \lt.
     - J(\al,\be) \mw  
       \fr{\mH - \mg}{p^2 - \mw} \sn^2 \theta_W
        (B_0 + 2 B_1) [h^0 w^\pm] \rt].
\end{eqnarray}
\begin{eqnarray}
  && \!\!\!\!\!\!\!\!\!\!\!\! 
     F^{\rm (c)} = \fr{2}{16 \pi^2 v^2 \cos \theta_W} \nn\\
    && \!\!\!\!\!\!\! \times \lt(    
 \fr{1}{2} \lt[ \;\;\tilde{\Pi}_{H^\pm w^\mp}^{(2)} \times 
\fr{1}{p^2 - \mw} \sn^2 \theta_W \rt. \rt. \nn \\
&&  -  \sn^2 \theta_W 
 \lt\{  \sn \ab \fr{1}{\mH} \tilde{T}_H 
     + \cs \ab \fr{1}{\mh} \tilde{T}_h \rt\} \nn \\
&& +  \sn^2 \theta_W \fr{\mg}{p^2 - \mw} 
\lt\{   \sn \ab \fr{1}{\mH} \tilde{T}_H 
     + \cs \ab \fr{1}{\mh} \tilde{T}_h \rt\}  \nn \\
&& - \sn^2 \theta_W \fr{1}{p^2 - \mw}
\lt\{ \sn \ab \tilde{T}_H + \cs \ab \tilde{T}_h \rt\} \nn \\
&& - \lt. \lt. \sn^2 \theta_W \fr{\mw}{p^2 - \mw} 
\lt\{   \sn \ab \fr{1}{\mH} \tilde{T}_H 
     + \cs \ab \fr{1}{\mh} \tilde{T}_h \rt\} \rt] \rt) \nn\\
&&\!\!\!\!\!\!\!\!\!\!\!\!\!\!
 =    \fr{\sin^2 \theta_W}{16 \mg v^2 \cos \theta_W}  
           \fr{1}{p^2 - m_W^2}\nn \\ 
&&\!\!\!\!\!\!\!\!\!\!\!\!\!
  \times \lt[ \lt\{ K(\al,\be) \mH - 2 J(\al,\be) \mH 
             - \fr{\sin (\al + \be)}
                  {\sin \be \cos \be} m_3^2 \rt\}  A(\mg)\rt. \nn\\
&&          + \lt\{ L(\al,\be) \mh + 2 J(\al,\be) \mh 
             - \fr{\cos (\al + \be)}
                  {\sin \be \cos \be} m_3^2 \rt\} A(\mg)  \nn\\
&& \!\!\!\!\!\!\!
       - \lt\{ K(\al,\be) \mH
                - \fr{\sin (\al + \be)}
                  {\sin \be \cos \be} m_3^2 \rt\} 
                  A(\mH)   \nn\\ 
&& \lt.-  \lt\{ L(\al,\be) \mh 
             - \fr{\cos (\al + \be)}
                  {\sin \be \cos \be} m_3^2\rt\} A(\mh) 
    + J(\al,\be) (\mH - \mh) A(\mw) \rt], 
\end{eqnarray}
where $\tilde{T}_H$ and $\tilde{T}_h$  are the tadpole graphs  
factorized by $1/(16 \pi^2 v^3)$. 
$\tilde{\Pi}^{(2)}_{H^\pm w^mp}$ is the $1/(16 \pi^2 v^2)$-factorized 
contribution of the two-point function 
which can be written in terms of the $A$-function.  
The full expressions for these are given in Eqs \eq{tad1}, \eq{tad2} 
and  \eq{two}.
\begin{eqnarray}
  && \!\!\!\!\!\!\!\!\!\!\!\! 
G^{\rm (a)} = \fr{2 \mw}{16 \pi^2 v^2 \cos \theta_W} \nn \\ 
&& \!\!\!\!\!\!\!\!\!\!\!\! 
     \times \lt[- \lt\{ K(\al,\be) \mH  + 
             J(\al,\be) (- \mH + 2 \mg) 
             - \fr{\sin (\al + \be)}
                  {\sin \be \cos \be} m_3^2\rt\} 
      (C_{12} + C_{23})[H^\pm A^0H^0] \rt.\nn \\
&& \!\!\!\!\!\!\!\!\!\!\!\! 
     - \lt\{ L(\al,\be) \mh  - 
             J(\al,\be) (- \mh + 2 \mg) 
             - \fr{\cos (\al + \be)}
                  {\sin \be \cos \be} m_3^2\rt\} 
      (C_{12} + C_{23})[H^\pm A^0h^0] \nn \\
&& \!\!\!\!\!\!\!\!\!\!\!\! 
     + \lt\{ K(\al,\be) \mH  + 
             J(\al,\be) (- \mH + 2 \mg) 
             - \fr{\sin (\al + \be)}
                  {\sin \be \cos \be} m_3^2\rt\}  
           \cs 2\theta_W (C_{12} + C_{23})[H^0H^\pm H^\pm] \nn \\
&& \!\!\!\!\!\!\!\!\!\!\!\! 
     + \lt\{ L(\al,\be) \mh  - 
             J(\al,\be) (- \mh + 2 \mg) 
             - \fr{\cos (\al + \be)}
                  {\sin \be \cos \be} m_3^2\rt\}  
           \cs 2\theta_W (C_{12} + C_{23})[h^0H^\pm H^\pm] \nn \\
&& \!\!\!\!\!\!\!\!\!\!\!\! 
     + J(\al,\be) (\mg - \mH)(C_{12} + C_{23}) [w^\pm z^0 H^0]\nn \\
&& \!\!\!\!\!\!\!\!\!\!\!\! 
     - J(\al,\be) (\mg - \mh)(C_{12} + C_{23})[w^\pm z^0 h^0] \nn \\
&& \!\!\!\!\!\!\!\!\!\!\!\! 
     - J(\al,\be) (\mg - \mH) \cs 2 \theta_W 
       (C_{12} + C_{23})  [H^0 w^\pm w^\pm] \nn \\
&& \!\!\!\!\!\!\!\!\!\!\!\! 
     + J(\al,\be) (\mg - \mh) \cs 2 \theta_W 
       (C_{12} + C_{23})  [h^0 w^\pm w^\pm] \nn \\
&& \!\!\!\!\!\!\!\!\!\!\!\!  
     - J(\al,\be) (\mg - \ma) 
      \lt\{ (C_{12} + C_{23}) [w^\pm H^0 A^0] - 
           (C_{12} + C_{23}) [w^\pm h^0 A^0] \rt\} \nn \\
&& \!\!\!\!\!\!\!\!\!\!\!\!  
     - J(\al,\be) \mw 
      \lt( 2 C_0 + 2 C_{11} + C_{12} + C_{23} \rt)[W^\pm H^0 A^0] \nn \\
&& \!\!\!\!\!\!\!\!\!\!\!\!  
     + J(\al, \be) \mw
     \lt( 2 C_0 + 2 C_{11} + C_{12} + C_{23} \rt)[W^\pm h^0 A^0]  \nn \\
&& \!\!\!\!\!\!\!\!\!\!\!\!  
     - J(\al,\be) \fr{\cs 2\theta_W}{\cs \theta_W} 
       m_W^2 (- C_{12} + C_{23}) [H^\pm H^0 Z^0] \nn \\
&& \!\!\!\!\!\!\!\!\!\!\!\! 
     + J(\al,\be) \fr{\cs 2\theta_W}{\cs \theta_W} 
       m_W^2  (- C_{12} + C_{23}) [H^\pm h^0 Z^0] \nn \\
&& \!\!\!\!\!\!\!\!\!\!\!\!  
     + J(\al,\be) \mw
     \lt( 2 C_0 - 2 C_{11} + 5 C_{12} + C_{23} \rt)[W^\pm Z^0 H^0]\nn \\
&& \!\!\!\!\!\!\!\!\!\!\!\!  
     - J(\al,\be) \mw
     \lt( 2 C_0 - 2 C_{11} + 5 C_{12} + C_{23} \rt)[W^\pm Z^0 h^0] \nn\\
&& \!\!\!\!\!\!\!\!\!\!\!\!  
     + J(\al,\be) \cs^2 \theta_W \mw 
     \lt( 4 C_{11} - 3 C_{12} - C_{23} \rt)[H^0 W^\pm W^\pm] \nn\\
&& \!\!\!\!\!\!\!\!\!\!\!\!  
     - J(\al,\be) \cs^2 \theta_W \mw 
     \lt( 4 C_{11} - 3 C_{12} - C_{23} \rt)[h^0 W^\pm W^\pm] \nn\\
&& \!\!\!\!\!\!\!\!\!\!\!\! \lt.
     + J(\al,\be) \mw \sn^2 \theta_W 
       \lt\{  \lt( C_{23} - C_{12} \rt)[H^0 w^\pm W^\pm] 
            - \lt( C_{23} - C_{12} \rt) [h^0 w^\pm W^\pm] \rt\} 
      \rt].
\end{eqnarray}
\begin{eqnarray}
G^{\rm (b)} = G^{\rm (c)} = H^{\rm (a,b \;{\rm and }\;c)} = 0.
\end{eqnarray}

The tadpole graphs $T_H (= \tilde{T}_H/(16 \pi^2 v^3))$ and 
$T_h (= \tilde{T}_h/(16\pi^2 v^3))$ are calculated as  
\begin{eqnarray}
T_H &=& \fr{1}{16 \pi^2 v^3}
\lt[  \mH \cs \ab \lt( A[w^\pm] + \fr{1}{2}A[z^0] \rt) \rt.\nn\\
&& + \lt\{ \mH \lt( \fr{\cs \al \sn^2 \be}{\cs \be} - 
                                   \fr{\sn \al \cs^2 \be}{\sn \be} \rt)
                 + 2 \mg \cs \ab 
             + \fr{\sin (\al + \be)}
                  {\sin \be \cos \be} m_3^2\rt\} A[H^\pm]  \nn \\
&&      + \lt\{ \mH \lt( \fr{\cs \al \sn^2 \be}{\cs \be} - 
                                   \fr{\sn \al \cs^2 \be}{\sn \be}
                                   \rt)
                 + 2 {\ma} \cs \ab   
             - \fr{\sin (\al + \be)}
                  {\sin \be \cos \be} m_3^2\rt\} \fr{A[A^0]}{2} \nn \\ 
&&      + \fr{3}{2} \lt\{ 
                         \lt( \fr{\cs^3 \al}{\cs \be} + 
                              \fr{\sn^3 \al}{\sn \be} \rt) \mH 
        - \fr{\cos 2 \be}{\cos \be \sin \be}\sin (\al - \be) m_3^2 
                    \rt\}  A[H^0] \nn\\ 
&&      + \lt\{ \fr{1}{2} ( \mH + 2 \mh ) \fr{\sn 2\al}{\sn 2\be}  
                - \fr{m_3^2}{4 \cos \be \sin \be}  
                (-3 \sin 2 \al + \sin 2 \be) \rt\}
                                           \cos (\al - \be) A[h^0] \nn \\
&&   \lt.   + 8 \cs \ab  \lt( \mw A[W^\pm] + 
                          \fr{1}{2} \mz A[Z^0] \rt) \rt] ,  \label{tad1}\\
    T_h &=&  \fr{1}{16 \pi^2 v^3} 
\lt[ - \mh \sn \ab \lt( A[w^\pm] + \fr{1}{2}A[z^0] \rt) \rt.\nn\\
&& + \lt\{\mh \lt( \fr{\sn \al \sn^2 \be}{\cs \be} - 
                                   \fr{\cs \al \cs^2 \be}{\sn \be} \rt)
                 - 2 \mg \sn \ab  
             + \fr{\cos (\al + \be)}
                  {\sin \be \cos \be} m_3^2 \rt\} A[H^\pm]  \nn \\
&& + \lt\{ \mh \lt( \fr{\sn \al \sn^2 \be}{\cs \be} - 
                                   \fr{\cs \al \cs^2 \be}{\sn \be} \rt)
                 - 2 \ma \sn \ab 
             + \fr{\cos (\al + \be)}
                  {\sin \be \cos \be} m_3^2\rt\} \fr{A[A^0]}{2} \nn \\ 
&& - \fr{3}{2} \lt\{ \lt( \fr{\sn^3 \al}{\cs \be} - 
                              \fr{\cs^3 \al}{\sn \be} \rt) \mh
        +\fr{\cos 2 \be}{\cos \be \sin \be}\cos (\al - \be) m_3^2 
                    \rt\} A[h^0]  \nn\\
&& + \fr{1}{2} \lt\{ ( 2 \mH + \mh ) \fr{\sn 2\al}{\sn 2\be} 
               - \fr{m_3^2}{4 \cos \be \sin \be}  
                (3 \sin 2 \al + \sin 2 \be) \rt\}\sn \ab A[H^0] \nn \\
&& \lt. - 8 \sn \ab \lt( \mw A[W^\pm] + \fr{1}{2} \mz A [Z^0] \rt) \rt] .  
\label{tad2}
\end{eqnarray}
Finally, $\Pi_{H^\pm w^\mp} (= \tilde{\Pi}_{H^\pm w^\mp}/(16 \pi^2 v^2))$ 
is given by 
\begin{eqnarray}
  \Pi_{H^\pm w^\mp}^{(2)}&=& \fr{1}{16 \pi^2 v^2}
      \lt[ 2(\mH - \mh) J(\al,\be) 
      \lt( A[W^\pm] + \fr{1}{4} A[Z^0]\rt) \rt. \nn \\
&& + 2 \lt\{ K(\al,\be) - J (\al,\be) 
             - \fr{\sin (\al + \be)}
                  {\sin \be \cos \be} m_3^2\rt\} \mH A[H^\pm]   \nn\\
&& + 2 \lt\{ L(\al,\be) + J (\al,\be) 
             - \fr{\cos (\al + \be)}
                  {\sin \be \cos \be} m_3^2\rt\} \mh A[H^\pm]   \nn\\
&& + \fr{1}{4} 
       \lt\{ \sn 2\be \lt(  \fr{\sn^2 \al}{\sn^2 \be} 
                        - \fr{\cs^2 \al}{\cs^2 \be}\rt) - 
           \sn 2 (\al - \be) \rt\}\mH A[A^0] \nn\\
&& + \fr{1}{4}
      \lt\{ \sn 2\be \lt(  \fr{\cs^2 \al}{\sn^2 \be} 
                        - \fr{\sn^2 \al}{\cs^2 \be}\rt) + 
           \sn 2 (\al - \be) \rt\}\mh A[A^0] \nn\\
&&- \cot 2 \be \;m_3^2\; A[A^0] \nn\\
&& + \fr{1}{4} \sn 2 \be 
     \lt( \fr{\sn^4 \al}{\sn^2 \be} - \fr{\cs^4 \al}{\cs^2 \be} 
          + \fr{\sn 2 \al \cs 2 \al}{\sn 2 \be} \rt) \mH A[H^0] \nn\\
&& + \fr{1}{4} \sn 2 \be 
     \lt(  \fr{\sn^2 \al \cs^2 \al}{\sn^2 \be} 
         - \fr{\sn^2 \al \cs^2 \al}{\cs^2 \be} 
         - \fr{\sn 2\al \cs 2 \al}{\sn 2\be} \rt) \mh A[H^0] \nn \\
&& - \fr{m_3^2}{2}\fr{\cos 2 \be}{\cos \be \sin \be} 
    \lt(\sin^2 (\al - \be) A[H^0] +  \cos^2 (\al - \be) A[h^0] \rt)\nn\\
&&   + J(\al,\be) \mg \lt( A[h^0] - A[H^0] \rt)  \nn \\
&& + \fr{1}{4} \sn 2 \be 
\lt( \fr{\sn^2 \al \cs^2\al}{\sn^2 \be} 
    - \fr{\cs^2 \al \sn^2 \al}{\cs^2 \be} 
    - \fr{\sn 2 \al}{\sn 2 \be} \cs 2 \al \rt) \mH A[h^0] \nn \\
&& \lt. + \fr{1}{4} \sn 2 \be 
\lt(  \fr{\cs^4 \al}{\sn^2 \al} - \fr{\sn^4 \al}{\cs^2 \be} 
     + \fr{\sn 2\al \cs 2\al}{\sn 2 \be}  \rt) \mh A[h^0] \rt] . 
\label{two}
\end{eqnarray}

\subsection*{APPENDIX 2:  Calculation of $H^+ \rightarrow W^+_L Z^0_L$ 
 by the use of the equivalence theorem}
\hspace*{18pt} 
As shown in Sec 4-1, at large $m_{H^\pm}$ region, the longitudinally 
polarized final gauge bosons become the dominant mode in 
$H^+ \rightarrow W^+Z^0$. 
In such case, we can check the results of the full calculation 
by the use of the equivalence theorem \cite{et0,et}, which says that 
$\Gamma(H^+ \rightarrow W^+_L Z^0_L) \sim 
 \Gamma(H^+ \rightarrow w^+   z^0)$ for $m_{H^\pm} \gg m_W$.      
The much simpler calculation of $\Gamma(H^+ \rightarrow w^+ z^0)$ 
can be useful to check the consistency of the full calculation. 
Here we show the explicit results of the amplitude 
${\cal M}_{ET}(H^+ \rightarrow w^+ z^0)$ calculated in the Landau gauge 
(See Fig 3.). 
We can clearly see in Eq \eq{amet} that ${\cal M}_{ET} \sim 0$ for 
$m_{H_\pm} \sim m_{A^0}$. 
\begin{eqnarray}
&& \!\!\!\!\!\!\!\! \!\!\!\!\!\!\!\!\!
{\cal M}_{ET}(H^+ \rightarrow w^+ z^0)=\fr{i}{v^3}(\mg - \ma) \nn\\
&& \!\!\!\!\!\!\!\!\!\!\!\! \times \lt[
\lt\{ K(\al,\be) \mH - J(\al,\be) (\mH - 2 \mg)   \rt\} 
 (\mH - \ma) C_0 [H^\pm,A^0,H^0] \rt. \nn\\
&& \!\!\!\!\! 
 + \lt\{ L(\al,\be) + J (\al,\be)(\mh - 2 \mg)  \rt\} 
 (\mH - \ma) C_0 [H^\pm,A^0,h^0] \nn\\
&& \!\!\!\!\!+ J (\al,\be) \mH (\mH - \ma) C_0[w^\pm,A^0,H^0]\nn\\
&&\!\!\!\!\! - J (\al,\be) \mh (\mh - \ma) C_0[w^\pm,A^0,h^0] \nn\\
&&\!\!\!\!\! - \lt\{ K (\al,\be) \mH - J (\al,\be) (\mH - 2 \mg) \rt\}
 B_0[H^0,H^\pm]\nn\\
&& \!\!\!\!\!- \lt\{ L (\al,\be) \mh + J (\al,\be) (\mh - 2 \mg) \rt\} 
B_0[h^0,H^\pm]\nn\\
&& \!\!\!\!\!+ J (\al,\be) (\mH - \mh) B_0[A^0,w^\pm]\nn\\
&& \!\!\!\!\!\!\!\!\!\! 
+ \lt\{  J(\al,\be) (\mh - \mH) + K (\al,\be) \mH 
                                   - L (\al,\be) \mh \rt\}
B_0[0;H^\pm,A^0]  \nn\\
&& \!\!\!\!\!+ J(\al,\be) \lt\{ \mh B_0[0;h^0,w^\pm] + \mh B_0[0;h^0,z^0]  
- \mH B_0[0;H^0,w^\pm] \rt. \nn\\ 
&& \!\!\!\!\!\!\!\!\!\!\!\!\!\!
\lt.\lt.- \mH B_0[0;h^0,z^0]
+  (\mH - \ma) C_0[0;H^0,A^0] - (\mh - \ma) B_0[0;h^0,A^0]\rt\} 
\rt]. \label{amet}
\end{eqnarray}

\newpage


\begin{thebibliography}{99}
\bibitem{lep} The LEP Collaborations, ALEPH, DELPHI, L3, OPAL,
              the LEP Electroweak Working Group and 
              the SLD Heavy Flavour Group, preprint CERN-PRE/96-183.
\bibitem{hmh} K. Hagiwara, S. Matsumoto, and D. Haidt, 
            Preprint KEK-TH-512, KEK Preprint 97-86, 
            DESY 96-192, {\tt  hep-ph/9706331}; 
            K. Hagiwara, S. Matsumoto, D. Haidt and C.S. Kim,
            Z. Phys. {\bf C64}, 559 (1994).              
\bibitem{lep2} A.Ballestrero {\it et. al.}, 
           {\it Proceedings of the Workshop on Physics at LEP2},
               G. Altarelli, T. Sj{\"o}strand and F. Zwirner (eds.), 
           CERN Yellow Report CERN 96-01 (1996). 
\bibitem{lhc} CMS Thechnical Proposal, CERN/LHCC/94-38,\\ 
              ATRAS Thechnical Proposal, CERN/LHCC/94-38.
\bibitem{lc} {\it Physics and Technology of the Next Linear Collider: 
                a Report submitted to Snowmass 1996,} 
             BNL 52-502, FNAL-PUB-96/112, LBNL-PUB-5425, SLAC Report 485, 
             UCRL-ID-124160;  {\it JLC-1,}KEK Report 92-16 (1992). 
\bibitem{hhg} J.F. Gunion, H.E. Haber, G. Kane and S. Dawson,
            {\it The Higgs Hunter's Guide}, 
             (Addison-Wesley, New York, 1990).
\bibitem{cleo}M.S. Alam et al. Phys. Rev. Let. {\bf 74}, 2885 (1995).
\bibitem{goto} T. Goto and Y. Okada, Prog. Theor. Phys. {\bf 94}, 407 (1995).
\bibitem{glme} J.A. Grifols and A. M{\'e}ndez, 
                    Phys. Rev. D {\bf 22}, 1725 (1980).
\bibitem{mepo} A. M{\'e}ndez and A. Pomarol, 
                    Nucl. Phys. {\bf B349}, 369 (1991). 
\bibitem{capd} M. Capdequi Peyran\`{e}re, H.E. Haber and P. Irulegui, 
               Phys. Rev. D {\bf 44}, 191 (1991). 
\bibitem{top}  CDF Collaboration, F. Abe {\it et al.}, 
                    Phys. Rev. Lett. {\bf 73}, 225 (1994).
\bibitem{4gen} T. Inami, T. Kawakami, and C.S. Lim, 
                 Mod. Phys. Lett. {\bf A 10}, 1471 (1995).  
\bibitem{exo}  R.M. Godbole, B. Mukhopadhyaya and M. Nowakowski,
               Phys. Lett. {\bf B352} 388  (1995).
\bibitem{riz}  T.G. Rizzo, Mod. Phys. Lett. {\bf A 4}, 2757 (1989).
\bibitem{dec}  T. Appelquist and J. Carrazone, Phys. Rev. D {\bf 11},  
                 2856 (1975).
\bibitem{ab}   T. Appelquist and C. Bernard, Phys.Rev. D {\bf 22}, 200 (1980). 
\bibitem{nondec} P. Ciafaloni and D. Esprin, 
                  Preprint UAB-FT-405, UB-ECM-PF-96/23,\\ 
                  ({\tt hep-ph/9612383}).
\bibitem{kt} S. Kanemura and H-A. Tohyama, KEK Preprint 97-96, 
                  KEK-TH-528, OU-HET 271, ({\tt hep-ph/9707454}). 
\bibitem{sc}  
       M. Veltman, 
             Acta. Phys. Pol. {\bf B 8}, 475 (1977), 
             Phys. Lett. {\bf 70 B}, 253  (1977); 
       M. Einhorn and J. Wudka, 
            Phys. Rev. D {\bf 39}, 2758 (1989),
            {\it ibid.} {\bf 47} 5029 (1993).
\bibitem{haber}H.E. Haber and A. Pomarol,
                     Phys. Lett. {\bf B302}, 435 (1993).
\bibitem{kkt} S. Kanemura, T. Kubota, and H-A. Tohyama, 
                 Nucl. Phys. {\bf B483}, 111 (1997).
\bibitem{gr}A.K. Grant, Phys. Rev. D {\bf 51}, 207 (1995).     
\bibitem{uni} S. Kanemura, T. Kubota and E. Takasugi,
            Phys. Lett. {\bf B313},  155 (1993).
\bibitem{back}J.F. Gunion, G.L. Kane, and J. Wudka, 
              Nucl. Phys. {\bf B299},  231 (1988).
\bibitem{glwe} S. Glashow and S. Weinberg, 
            Phys. Rev. D {\bf  15}, 1958 (1977). 
\bibitem{cp} S. Weinberg, Phys. Rev. D {\bf 42}, 860  (1990). 
\bibitem{cus}  S. Weinberg, Phys. Rev. D {\bf 19}, 1277 (1979); 
               L. Susskind, {\it ibid.} {\bf 20}, 2619 (1979).
\bibitem{georgi} H. Georgi, Hadir. J. Phys. {\bf 1}, 155 (1978).
\bibitem{et0} 
       J.M. Cornwall, D.N. Levin, and G. Tiktopoulos, 
            Phys. Rev. D {\bf 10}, 1145 (1974);
       B.W. Lee, C. Quigg, and H.B. Thacker, 
            {\it ibid.} {\bf 16}, 1519 (1977).
\bibitem{et}  
   M.S. Chanowitz and M.K. Gaillard,
              Nucl. Phys. {\bf B261}, 379 (1985);
   H-J. He, Y-P. Kuang and X. Li,
              Phys. Rev. D {\bf 49}, 4842 (1994), and references therein.  
\bibitem{kiyo} J. Hisano, S. Kiyoura, and H. Murayama, 
                 Phys. Lett.{\bf B399}, 156 (1997). 
\bibitem{bb}   V.Barger, J.L.Hewett and R.J.N.Phillips, 
               Phys. Rev. D {\bf 41}, 3432 (1990); 
               A.J. Buras, P. Krawczyk, M.E. Lautenbacher, and
               C. Salazar, Nucl. Phys. {\bf B337}, 284 (1990). 
\bibitem{dom} J. Preskill, S.P. Trivedi, and F. Wilczek, 
                 Nucl. Phys. {\bf B363}, 207 (1991).            
\bibitem{e6}   T.G. Rizzo,  Phys. Rev. D {\bf 39}, 728 (1989). 
\bibitem{bhs}  R.M. Barnett, H.E. Haber, and D.E. Soper, 
               Nucl. Phys. {\bf B306}, 697 (1988). 
\bibitem{pave} G. Passarino and M. Veltman, Nucl. Phys. {\bf B160}, 151 (1979).
\end{thebibliography}
\end{document}